\documentclass[preprint,pre,showkeys,showpacs]{revtex4-1}

\usepackage{epsfig}
\usepackage{amsmath}

\begin{document}
\title{Competitive cluster growth on networks: complex dynamics and survival strategies}
\author{N.Nirmal Thyagu\footnote{Present Address: Dept. of Biomedical Engineering, Rutgers University, Piscataway, NJ 08854}}
\email{nnirmal@soemail.rutgers.edu}
\author{Anita Mehta}
\email{anita@bose.res.in}
\affiliation{Department of Theoretical Studies, S. N. Bose National Centre for Basic Sciences, Sector-III, Block-JD, Salt Lake, Kolkata-700078, India}
\date{\today}
\begin{abstract}
We extend the study of a model of competitive cluster growth in an active medium from a regular topology to a complex network topology;
 this is done by  adding nonlocal connections with probability
 $p$ to sites on a regular lattice, thus enabling one to interpolate between regularity and full randomness.
 The model on networks demonstrates high sensitivity to small changes in initial configurations,
 which we characterize using  damage spreading. The main focus of this paper is, however,
 the devising of survival strategies through selective networking, to alter the fate of an arbitrarily chosen cluster: whether this
be to revive a dying cluster to life, or to make a weak survivor into a stronger one. Although
such goals are typically achieved by networking with relatively small clusters,
 our results suggest that it ought to be possible also to network successfully with peers and larger clusters. The main indication
of this comes from  the probability distributions of mass differences between survivors and their immediate
neighbours, which  show an interesting universality; they suggest strategies for winning against the odds. 
\end{abstract}

\maketitle

\section{Introduction}

The study of complex systems with interacting agents is of great
 interest in very many fields  ranging from physics through quantitative
 biology to economics \cite{blasius}. Modelling the  dynamics of 
 interacting species should capture the competition between the species 
 for growth and survival, which underlies all evolutionary dynamics.
 One of the most famous  models of this class is the
 Lotka-Volterra model used for studying predator-prey interactions
 \cite{takeuchi}, which is loosely predicated on the principle of the survival
 of the fittest \cite{darwin}.

 Typically, little attention has been paid to the case where agents interact
 through an active medium; this is a case where the
  Lotka-Volterra model cannot be simply applied.  
 A very recent study \cite{massie_pnas2010} on microbial populations strongly suggests that the interactions
between the individuals in the population occur through an active nutrient pool.
For instance, in the case of interacting traders, a central exchange or reserve bank could represent such
 an active medium; this could selectively deplete or enrich individuals
 depending on taxation modes in the system. Our present study is based on such
 an example \cite{am_wealth}; the detailed model of competing clusters on which it
 is based \cite{luck_am},  forms the basis of  our study here. 
The term `cluster' is used throughout this
paper to refer a single coarse grained macroscopic unit that is a collection of numerous small microscopic subunits. The magnitude 
of such a cluster is characterized by its absolute mass.

 Beginning  with the appearance of two seminal papers \cite{watts,albert}, the study of complex networks has attracted phenomenal attention from diverse fields. The spur in the growth of these studies began primarily with the prospect that numerous systems in nature can be compartmentalized into  highly interconnected entities of dynamical units. To quote a few of the numerous examples that have a complex network topology  : the Internet, social interactions, biological metabolic networks, predator-prey interactions in food chains, etc (see \cite{rev_comnets}). Many real world networks, in spite of their inherent differences,  have been found to have complex network topology, for example : social \cite{soc_nets} and biological networks \cite{bio_nets}.

Predator-prey interactions seen in food webs and ecological networks are of special importance to our present study. Of 
the numerous studies done on food webs, some of them have a small world structure \cite{montoya,dunne_02}, while some other  studies suggest that the food webs do not show either a small world or  a scale-free topology but may have a different network structure \cite{garlaschelli}. 
While the exact form of the structure of a predator-prey system is subjective, the nature of the interaction has been shown to be highly nonlinear and therefore can be modelled mathematically \cite{fussmann2000}.

In the light of the above studies, it is  becoming increasingly clear that 
interactions between agents are neither of the mean field type, where all of the species interact with all others,  nor necessarily confined to  local neighbourhoods. Therefore complex networks are a far more appropriate basis for the modelling of most systems with their combination of globality and locality, as well as their freer choice of agent-agent contacts. One of the main aims of this paper is thus to study the competitive cluster model of  \cite{luck_am} on networks.

The plan of the paper is as follows. First, we introduce the model (Sec. \ref{sec_model}); next, we do a deeper investigation of the problem in two dimensions in Sec. \ref{regular}.  Here, we demonstrate the inherent complexity of the problem in finite-dimensional systems, where, contrary to the mean-field case, the largest cluster is not necessarily a survivor. Next, we investigate the main features of the model on a network. Starting with a regular network (which here is a finite-dimensional lattice) with nearest-neighbour interactions, we add non-local links to the existing lattice, with an associated probability $p$, in the range $0 \leq p \leq 1$ \cite{watts}.  We measure the survivor ratios as a function of the wiring probability $p$, as $p$ is increased to reflect the topologies of small world networks and fully random networks  ($p=1$). Our system turns out to exhibit great sensitivity to small  changes in the initial distribution of clusters; we characterize this using the concept of damage spreading (Sec. \ref{sec_dam}).

In Sec. \ref{networksmall}, we ask the following question: can the destiny of a selected cluster be changed by networking? By `networking' we mean there is a two way interaction between two or more clusters. We evolve a  strategy to do this by  selectively networking a cluster of our choice with non-local clusters based on their initial we
masses. One of our most interesting observations in this paper is that  a cluster which would die in its original 
neighbourhood, \textit{is indeed able to change its fate}, becoming a survivor via selective networking.
 Next, in Sec. \ref{univ} we calculate the pair-wise  probability for a cluster to survive against larger and smaller neighbours; the probability distributions obtained show an interesting universality, independent of the mass considered.  We discuss the implications of these and other results in our concluding section, Sec. \ref{discuss}.

\section{Model\label{sec_model}}

The  present model was first used in the context of cosmology to describe the mass accretion of  black holes in the presence of a radiation field \cite{archan}. Its applications, however, are considerably more general; used in the context of economics \cite{am_wealth}, it manifested an interesting rich-get-richer behaviour. Here, we recapitulate some of its principal properties \cite{luck_am}.

Consider an array of immobile clusters with time-dependent masses $m_i (t)$ located at equi-spaced regular lattice sites. The time evolution of masses of the clusters is given by the coupled deterministic first order equations,
\begin{eqnarray}
\frac{dm_i}{dt} &=& \left(\frac{\alpha}{t} - \frac{1}{t^{1/2}} \sum_{j \neq i} g_{ij} \frac{dm_j}{dt} \right) m_i  - \frac{1}{m_i}.
\label{blackhole}
\end{eqnarray}
Here, the parameter $\alpha$ is called the mass accretion parameter  and $g_{ij}$ defines the coupling between the clusters $m_i$ and $m_j$; there is no self coupling.
The first term in the R. H. S of  Eqn. \ref{blackhole} represents the gain term, which is composed of two terms - the free rate of growth of an individual cluster (proportional to the parameter $\alpha>1/2$), and the rate of growth induced by all other clusters coupled through the surrounding medium.  The coupling $g_{ij}$ between two clusters is proportional to the inverse square distance ${d_{ij}}^2$ between them at an initial time $t_0$ \cite{luck_am}. The negative second term, $-1/m_i$, represents
the dissipation to the surrounding medium.

A logarithmic time is introduced in the  study for convenience. We define a reduced time $s = \ln(t/t_0)$, where $t_0$ is the initial time. Similarly, for convenience, we define reduced masses $X_i = m_i/t^{1/2}$. Using the new variables, Eqn . \ref{blackhole} can be rewritten as,
\begin{eqnarray}
\frac{dX_i}{ds} \equiv {X'}_i = \left(\frac{2 \alpha -1}{2} - \sum_{j \neq i} g_{ij} \left(\frac{X_j}{2} + {X'}_j  \right) \right) X_i  - \frac{1}{X_i}
\label{model}
\end{eqnarray}
The primes denote the differentiation performed with respect to the reduced time variable $s$ throughout.

\begin{figure}[!t]
\centering
\resizebox{90mm}{70mm}{\includegraphics{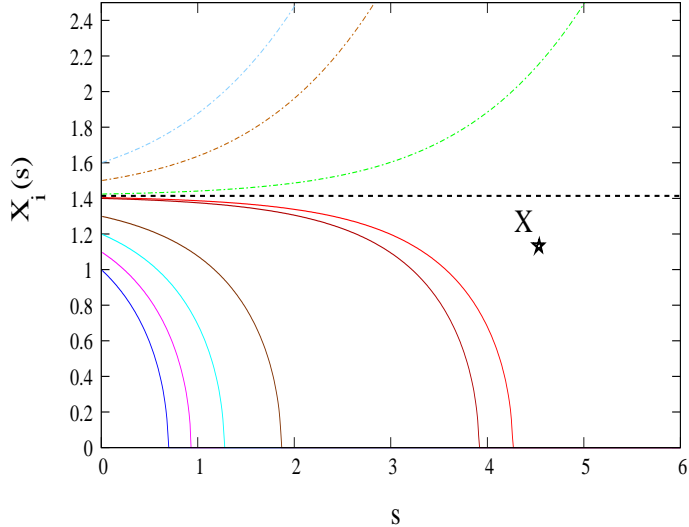}}
\caption{(color online) The plots show the evolution of individual non-interacting clusters with a range of initial masses obeying Eqn. \ref{model}. Clusters with initial mass $X_0$  greater than $X_{\star}$ live forever, or else they die in time. Here $\alpha = 1.0$, therefore $X_{\star} = \sqrt{\frac{2}{2 \alpha-1}} = \sqrt{2}$. \label{freemasses}}
\end{figure}

Continuing our recapitulation of the results of the model \cite{luck_am}, we consider a scenario where there is a single isolated cluster, whose initial reduced square mass is $X_0$. Under the dynamics defined by Eqn. \ref{model}, the cluster will be a survivor only if $X_0>X_{\star} (= \sqrt{\frac{2}{2 \alpha-1}})$, else it will eventually die out (see Fig. \ref{freemasses}) \cite{luck_am}. Next, consider a system of two clusters with equal  initial masses; it has been shown \cite{luck_am} that there exists a critical coupling $g_c$ such that for $g<g_c$ they both survive if their initial masses are greater than $X_{\star}$. In the case of  two clusters with unequal masses, the smallest of them dies first, in reduced time $s_1$, and the largest cluster will either survive (if $X(s_1) > X_{\star}$) or die (if $X(s_1) < X_{\star}$).

Consider the limit of an infinitely large number of clusters all connected to each other; this represents a limiting mean field regime, with fully collective behaviour involving long-range interactions.  For $g>g_c$, all the clusters but the largest will eventually die out. In the weak coupling regime ($g<g_c$), which is the  focus of  this paper, the dynamics consist of two successive stages \cite{luck_am}. In Stage I, the clusters behave as if they were isolated;  they grow (or die) quickly if their masses are greater (or less) than the threshold  $X_{\star}$. In Stage II, slow  collective dynamics leads to a scenario where, again, only the single largest cluster survives. This weakly interacting mean field regime shows the presence of two well-separated time scales, a characteristic feature of  glassy systems \cite{luck_am}. 

\begin{figure}[!t]
\begin{center}
\resizebox{90mm}{70mm}{\includegraphics{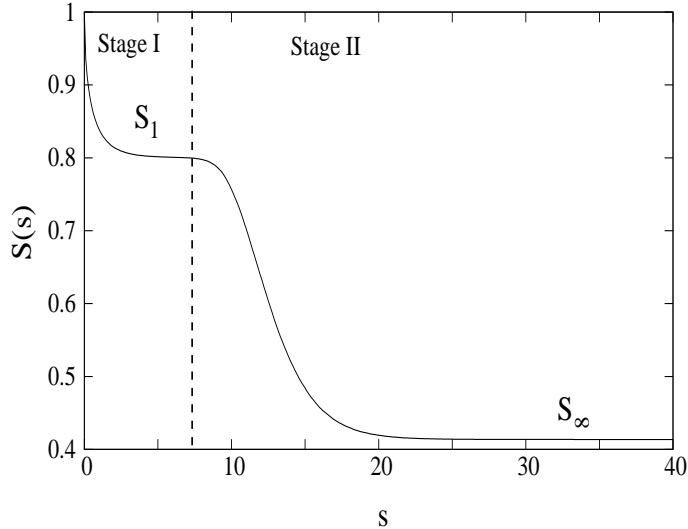}}
\end{center}
\caption{ The survival ratio $S(s)$ plotted as a function of reduced time $s$, for  clusters
 distributed in a regular one-dimensional lattice of size $100,000$. In Stage I, clusters grow
independently, while in Stage II the growth is collective. Here, $S_1 = 0.8$ is the survival
ratio at the end of Stage I and   $S_{\infty}$ is the asymptotic survival ratio. \label{surratio}}
\end{figure}

Similar glassy  dynamics  also arise  when the model is solved on a periodic lattice with 
nearest-neighbour interactions. The dynamical equations in  Eqn. \ref{model} take the form \cite{luck_am}:

\begin{eqnarray}
X'_{{\bf n}} &=& \left( \frac{2 \alpha - 1 }{2} + g \sum_{{\bf m}} \left( \frac{1}{X_{\bf m}} - \alpha X_{\bf m} \right) \right) X_{\bf n} - \frac{1}{X_{\bf n}},
\label{finite_model}
\end{eqnarray}
by keeping the terms upto first order in $g$.
Here ${\bf m}$ runs over the $z$ nearest neighbours of the site ${\bf n}$, where for a 
one-dimensional ring topology $z=d$, for a two-dimensional lattice $z=2d$, and so on for any
 space of dimension $d$. We summarize earlier results on the dynamics: 
 In Stage I, clusters evolve independently; only those whose initial mass $X_i (s=0)>X_{\star}$ 
survive this stage. In Stage II, the dynamics are slow and collective; the survivors
of this stage are fully isolated from each other by `dead' sites,  and
 their number asymptotes to  a constant $S_{\infty}$ (Fig. \ref{surratio}). 
These isolated clusters  survive forever.

Following the mean field scenario, where the largest cluster is the only one to survive, we ask
if this would also be the case for the finite-dimensional case: that is, 
are the survivors
 the largest in the locality? Somewhat surprisingly, this
is not always the case. While
 one can certainly rule out the survival of a cluster whose initial mass is less than threshold
  ($X_{\star}$),  many-body interactions give rise to extremely
 complex dynamics in Stage II for  clusters with  $X>X_{\star}$. The
 asymptotic configurations of survivors
 in the finite-dimensional case are thus
often nontrivial and counter-intuitive (see Sec. \ref{regular}). 

\section{Nontrivial dynamics in two-dimensional regular lattices \label{regular}}

In this section we do a deeper investigation of the cluster growth
 model \cite{luck_am}, with individual clusters  placed on the vertices of a
 regular two-dimensional lattice with  nearest-neighbour interactions (coordination 
number for all the clusters in the lattice is $z=4$).
While the results of Sec. \ref{sec_dam} will show the
 extreme sensitivity of the model to the distribution of cluster masses,
the results of this section will help us get a flavour of  the complexity displayed by the model.

We choose a $50 \times 50 $  regular lattice for our investigations in this 
paper, unless otherwise specified; 
we have checked that there are no observable finite size effects for this
choice.  Initial masses are characterized by an exponential
 distribution $P(X_0) = \mu \exp(- \mu X_0)$, 
where the mean is taken to be $\mu=-\log(S_{(1)})/X_{\star}$, 
for different values of $S_1$. 
The coupling parameter is set 
 at $g=0.001$ and the parameter $\alpha$ set to  $1.0$
 \cite{luck_am} .

\begin{figure}[!t]
\begin{center}
\begin{tabular}{cc}
\hspace{-1cm}(a)&
\hspace{-0.4cm}\resizebox{90mm}{45mm}{\includegraphics{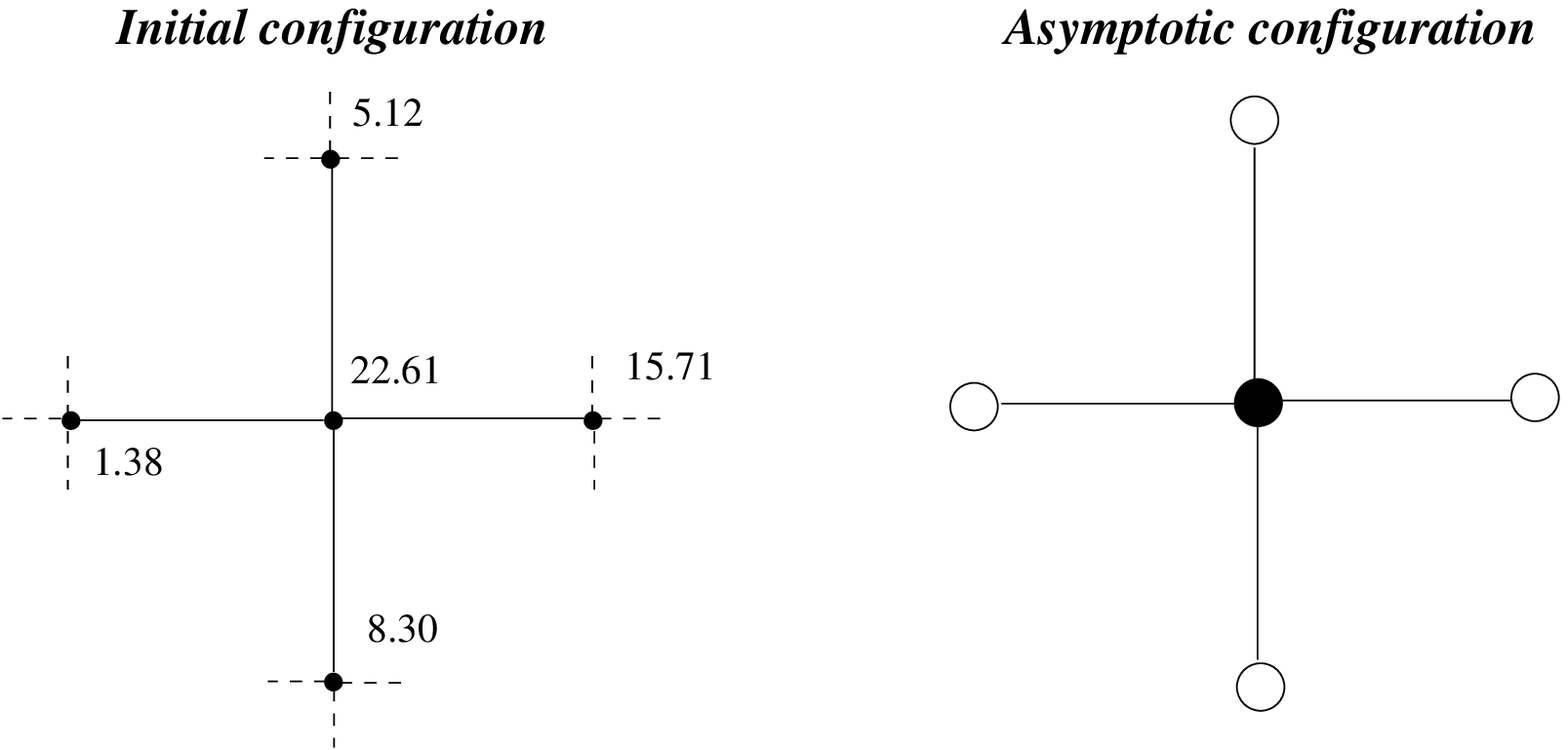}}\\
\hspace{-1cm}(b)&
\hspace{-0.4cm}\resizebox{90mm}{45mm}{\includegraphics{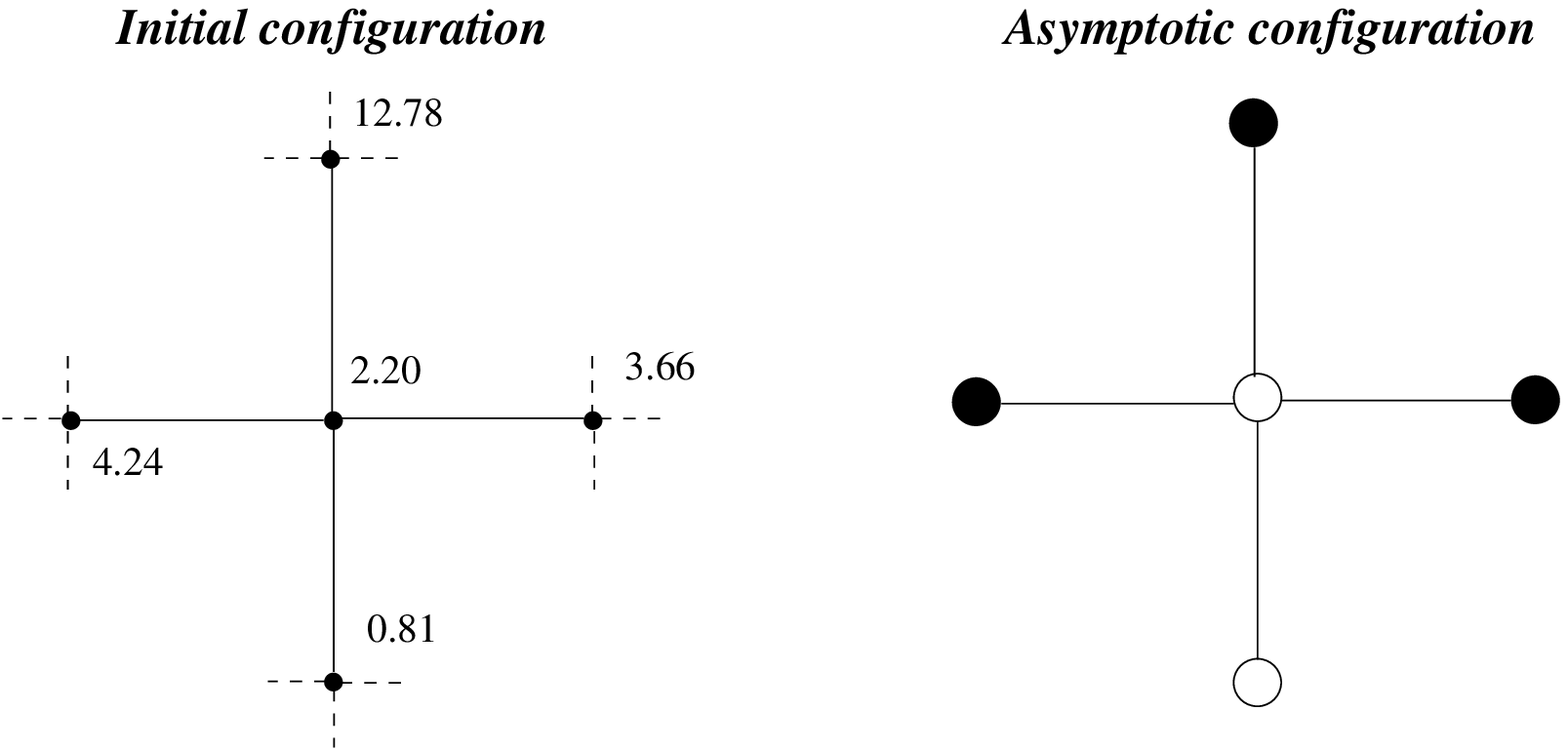}}\\
\end{tabular}
\end{center}
\caption{A sample configuration of a subset of the system is shown here. The left panels show the cluster configurations at time $s=0$. The right panels show the asymptotic configurations. (a) The central cluster dominates its immediate neighbours and emerges as the survivor. (b) Three larger neighbours eventually dominate the central cluster to become survivors by killing it. Open and filled circles represent non-survivors and survivors, respectively.  \label{cm_obvious}}
\end{figure}

First we consider two of the more obvious  cases for demonstration (Fig. \ref{cm_obvious}). In Fig. \ref{cm_obvious}(a), the central cluster has the
 largest mass among its neighbours, and in Fig. \ref{cm_obvious}(b),
 the central cluster has  the smallest mass of those of its three
 neighbours which are larger than threshold (the fourth one will obviously not
 survive beyond Stage I).

These two cases show results similar to the mean field scenario \cite{luck_am}; 
 the central cluster with the largest
 initial mass becomes a survivor (Fig. \ref{cm_obvious}(a))
 and the central cluster with the smallest initial  mass 
dies off (Fig. \ref{cm_obvious}(b)). 

\begin{figure}[!t]
\begin{center}
\hspace{-0.4cm}\resizebox{90mm}{90mm}{\includegraphics{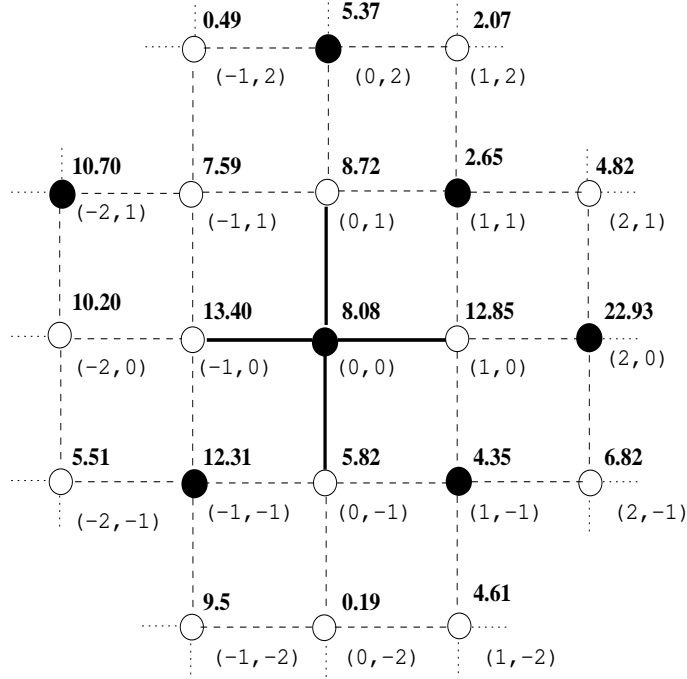}}\\
\end{center}
\caption{A small region of the two-dimensional lattice showing neighbouring
 clusters with comparable initial masses. The numbers show the initial masses; open and filled circles represent non-survivors and survivors, respectively, in the asymptotic limit. This is an illustration of the complex dynamics referred to in
the text.
\label{grid_complex}}
\end{figure}

However, in general, 
such simplistic reasoning may not work.
In a general scenario such as in Fig. \ref{grid_complex}, consider the 
cluster located at co-ordinates $(0,0)$, which has three larger and one
 smaller neighbours. Reasoning naively, we might expect that the larger
 neighbours $X_{(1,0)},X_{(-1,0)},X_{(0,1)}$ will eventually be able to kill
 the smaller central cluster  $X_{(0,0)}$. Contrary to this, we find that
 in the asymptotic limit, the central cluster  $(0,0)$ is a survivor.
 Similar results are obtained for the  clusters
  at the coordinates $(1,1)$ and $(1,-1)$; although surrounded by larger
 neighbours, they nevertheless survive against the odds.
Another apparently counter-intuitive case is illustrated in
 Fig. \ref{grid_complex} for the cluster with
coordinates $(-1,0)$. To start with, all the neighbours are smaller than the
 central cluster, but eventually the two neighbours at $(0,0)$ and $(-1,1)$
 survive at the expense of the central cluster, which dies.

The explanation for such apparent anomalies is the many-body nature
of this problem; the  evolution of neighbouring sites is  influenced by
 their  neighbours, and so on. 
Examining the dynamical equation given in Eq. \ref{finite_model},
we see that the overall  contribution of the ${\bf m}=4$ immediate neighbours
 of a given cluster at any  instant of time is expressed by the summation, $g \sum_{\bf m} \left( \frac{1}{X_{\bf m}} - \alpha X_{\bf m} \right)$, where $X_m$ denotes the reduced masses of the $m$ neighbours. This term should be positive for
  ${X'}_{\bf n}$ to be positive, which means that the overall  rate of growth
 of its neighbours  $\sum {X'}_{\bf m}$ should be negative.

 This can also be understood  by inspecting Eqn. \ref{model}. In Eqn. \ref{model}, if $\sum {X'}_{j}$ is positive, then the overall growth rate of the neighbours of a cluster $i$ under consideration is positive. Then the rate of growth of that cluster $i$, ${X'}_i$, will eventually go negative. On the other hand, a negative $\sum {X'}_{j}$ will increase the rate ${X'}_i$
and enhance its growth. The precise manner in which the $\sum {X'}_{j}$ evolves is complex given the many-body nature of the interactions (see Figs. \ref{636} and \ref{1054} for sample  time evolution plots of the combined rates of growth of the neighbouring clusters contrasted with the center cluster). Therefore, for a cluster to become a survivor, it should grow constantly;
 its rate of growth $X'_{\bf n}$ should thus  always be positive.

In other words: Clusters that survive  stage I will be the ones with $X>X_{\star}$ 
and  in stage I they grow independently; there is thus no reason for their rate to be negative.
On the other hand in stage II,  when there are interactions
between clusters, even if a cluster temporarily has a negative growth rate, it will be fed on by its competitors and killed off.

\begin{figure}[!t]
\begin{center}
\hspace{-0.4cm}\resizebox{120mm}{110mm}{\includegraphics{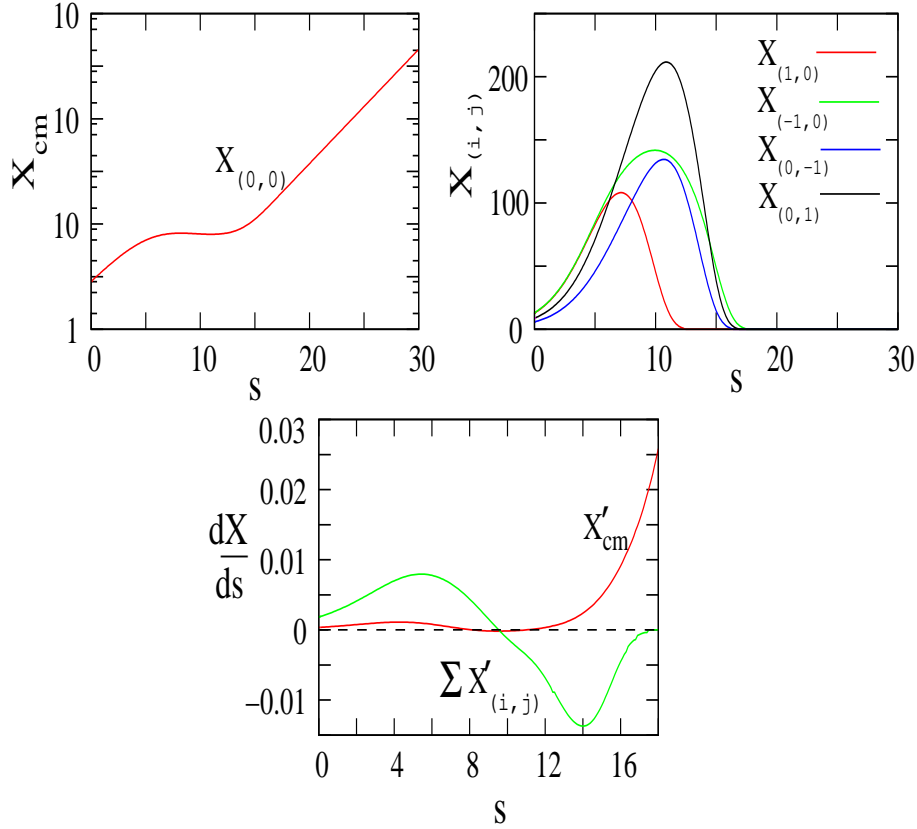}}\\
\end{center}
\caption{ (Color online) (a) The central cluster $X_{(0,0)}$ as a function of reduced time $s$. (b) The neighbouring clusters around $(0,0)$ as a function of $s$. (c) The rate of evolution of the central cluster and the combined rates of evolution of the immediate neighbours are compared - the central cluster wins the competition.\label{lat_evol1}}
\end{figure}

Next, we present plots showing the evolution of a given cluster and its
 immediate neighbours for the cases discussed in Fig. \ref{grid_complex}.
Consider the evolution of the cluster at  $(0,0)$,  shown in 
Fig. \ref{lat_evol1} (a). The immediate neighbours of this cluster are seen to
 die when the central cluster grows (see Fig. \ref{lat_evol1}(b)).
In Fig. \ref{lat_evol1}(c) we plot the rate of growth of the central cluster
 (in `red') and the overall growth of the immediate neighbours 
$\Sigma {X'}_{(i,j)}$ (in `green'). Initially, 
both ${X'}_{(0,0)}$ and $\Sigma {X'}_{(i,j)}$ are positive (Stage I), with 
the neighbours growing faster than the central site, in accord with intuitive
expectations  based on their initial masses. When 
the collective dynamics of Stage II sets in,  $\Sigma {X'}_{(i,j)}$ falls
 and becomes negative; this is a demonstration of
  many-body interactions in action, where neighbours are influenced by
 their neighbours, and so on.  Finally, soon after the neighbours begin
to decay ($\Sigma {X'}_{(i,j)}$ gets negative), the rate of growth of the central
site,  ${X'}_{(0,0)}$, shoots up, making it
 a survivor that continues to grow forever.

\begin{figure}[!t]
\begin{center}
\hspace{-0.4cm}\resizebox{120mm}{110mm}{\includegraphics{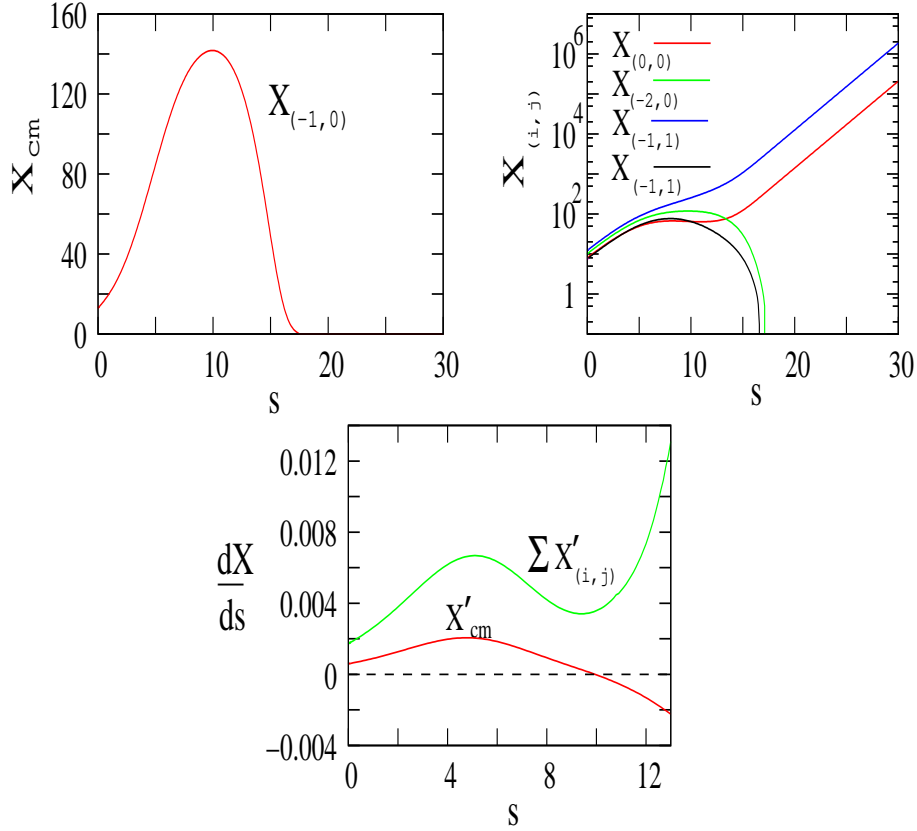}}\\
\end{center}
\caption{(Color online) (a) The growth of the central cluster $X_{(-1,0)}$ as a function of reduced time $s$. (b) The growth of the neighbouring clusters around $(-1,0)$ as a function of $s$. (c) The rate of evolution of the central cluster and the combined rates of evolution of the immediate neighbours are compared - 
the central site loses. \label{lat_evol2}}
\end{figure}

Graphs similar to those in Fig. \ref{lat_evol1} have been obtained
for the other survivors of  Fig. \ref{grid_complex}, viz. $X_{(1,1)}$ and $X_{(1,-1)}$; we have not plotted them here,
in the interests of brevity. To complete our analysis, we show however the
graphs in a  contrasting case, where a given cluster eventually dies, despite being
surrounded by smaller neighbours. Choosing $X_{(1,-1)}$ in 
Fig. \ref{grid_complex}, we show
 the growth of the central site and its associated neighbours, respectively,
 in Fig. \ref{lat_evol2} (a) and (b).
 Here, contrary to expectations, the central cluster dies in a finite time, 
while two of its neighbours manage to survive. 
 These dynamics are seen clearly in Fig. \ref{lat_evol2} (c); 
in the fast dynamics of Stage I, the combined rate of growth of the neighbours
 is faster than the central cluster. After the onset of slow dynamics in Stage
 II, the rate of growth of the central cluster goes negative,
 which contributes to the growth of its neighbours.  

Clearly from the above, the simple rule of the mean field scenario, that
 the largest cluster is always a survivor, does not hold for 
 lattice-based models with nearest-neighbour interactions. In fact, it
is rare to be able to find cases like
 Figs. \ref{cm_obvious} (a)-(b), where
 we can ignore the effect of further neighbours.
The generic case is much more as depicted in
 Fig. \ref{grid_complex}, where  information on the immediate neighbourhood
of a site does {\it not} provide a reliable indication of its fate, highlighting
the innate complexity of our model on a lattice.

\section{Cluster dynamics in complex networks\label{sec_compnet}}

In the previous section we discussed the complex dynamics exhibited by the cluster growth model on a lattice.  However, both lattices and mean field scenarios are topologically unrealistic - recent work indicates \cite{montoya, dunne_02, fussmann2000} that most real-world systems are more appropriately solved on networks, which are spatially disordered arrays with non-trivial connections between their sites. A particularly interesting example is the class of small world networks, which are constructed by starting with a regular lattice, and adding links randomly to its sites with  probability $p$  \cite{watts} . Alternative constructions are also possible, where existing links are `rewired';  i.e. existing links are severed and then reconnected to  randomly chosen lattice sites with probability $p$. Small world networks have the property that  long-range interactions can co-exist with short-range interactions; they contain realistic elements like `hubs', where certain sites are preferentially endowed with many connections, as is the case for example, in the aviation industry \cite{boccal}. 

 We shall in the following apply the former construction,  whereby new links are added randomly with a given probability. In addition, we shall freeze the newly added links for all times, rather than evolving them continuously - which we call  `static wiring' \cite{zah}. As  a result of this change, the average degree of the sites will be higher for all $p>0$.

\begin{figure}[!t]
\begin{center}
\begin{tabular}{cc}
\hspace{-0.4cm}\resizebox{70mm}{60mm}{\includegraphics{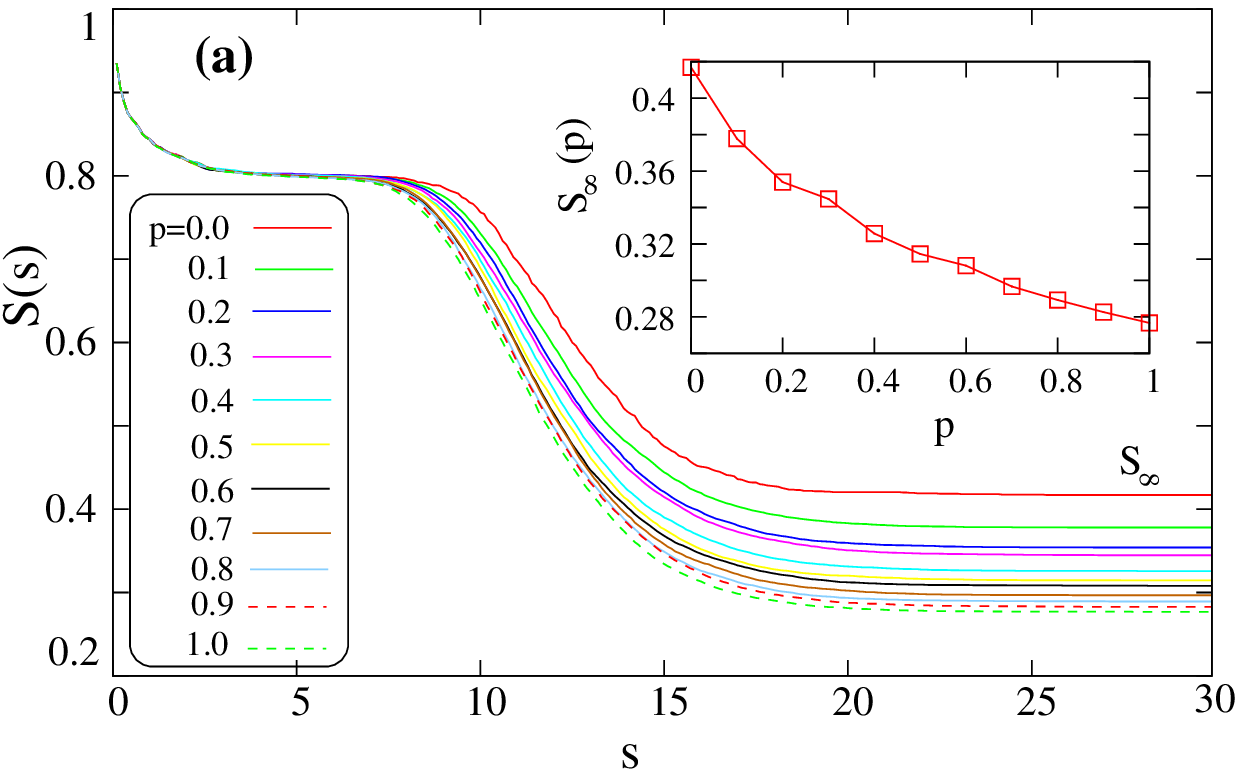}}&
\hspace{0cm}\resizebox{70mm}{60mm}{\includegraphics{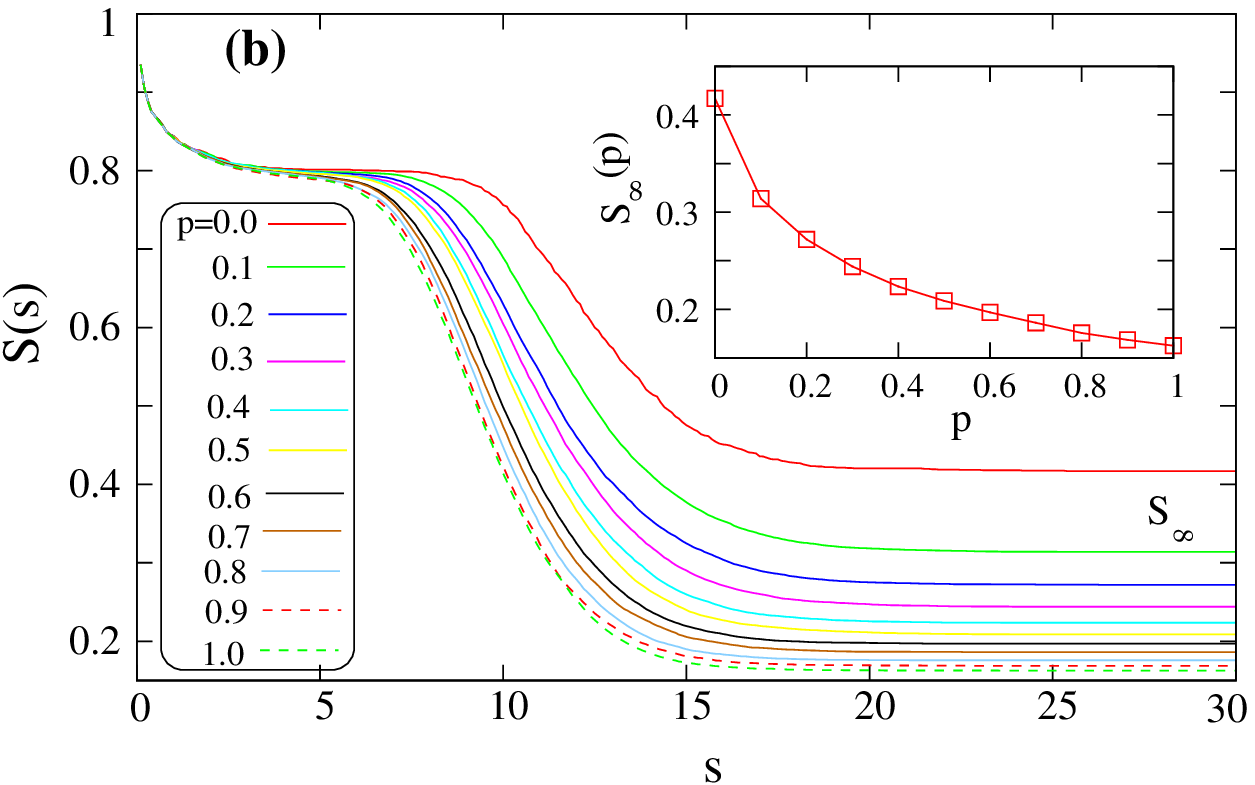}}\\
\end{tabular}

\begin{tabular}{cc}
\hspace{-0.4cm}\resizebox{70mm}{60mm}{\includegraphics{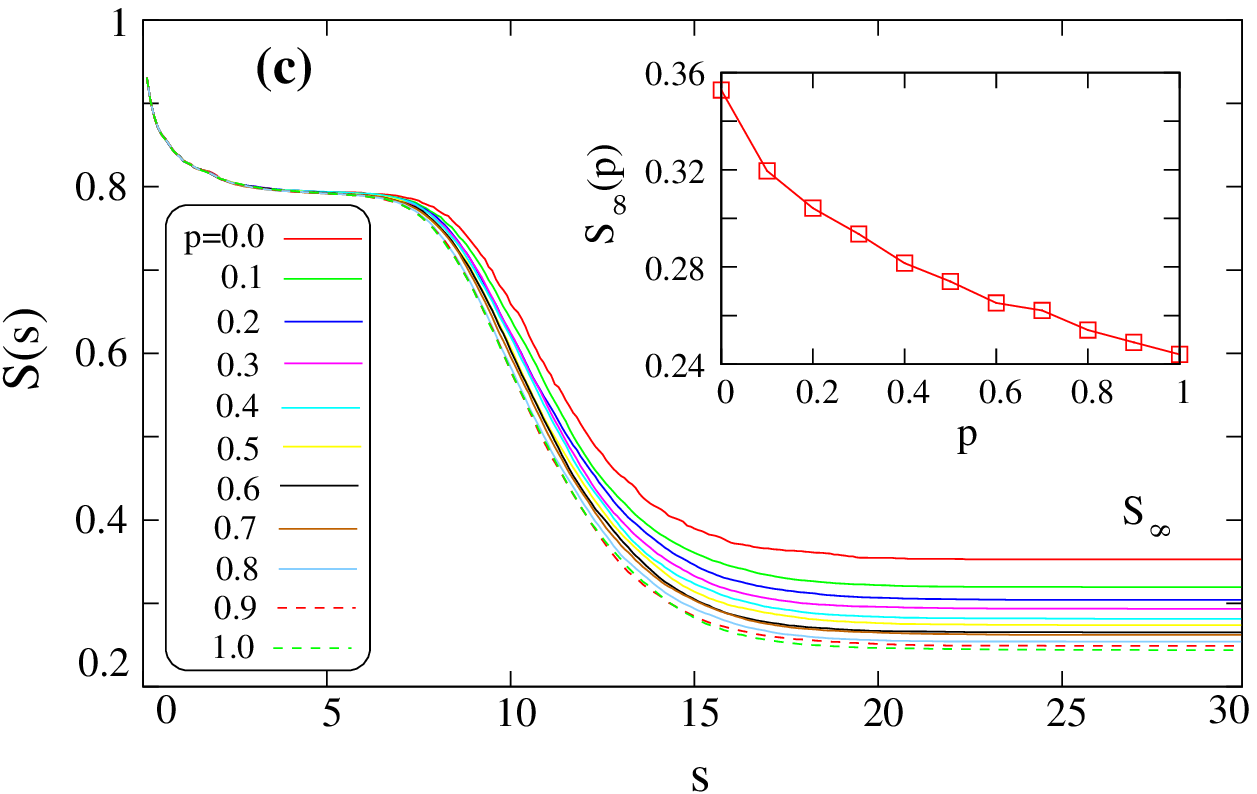}}&
\hspace{0cm}\resizebox{70mm}{60mm}{\includegraphics{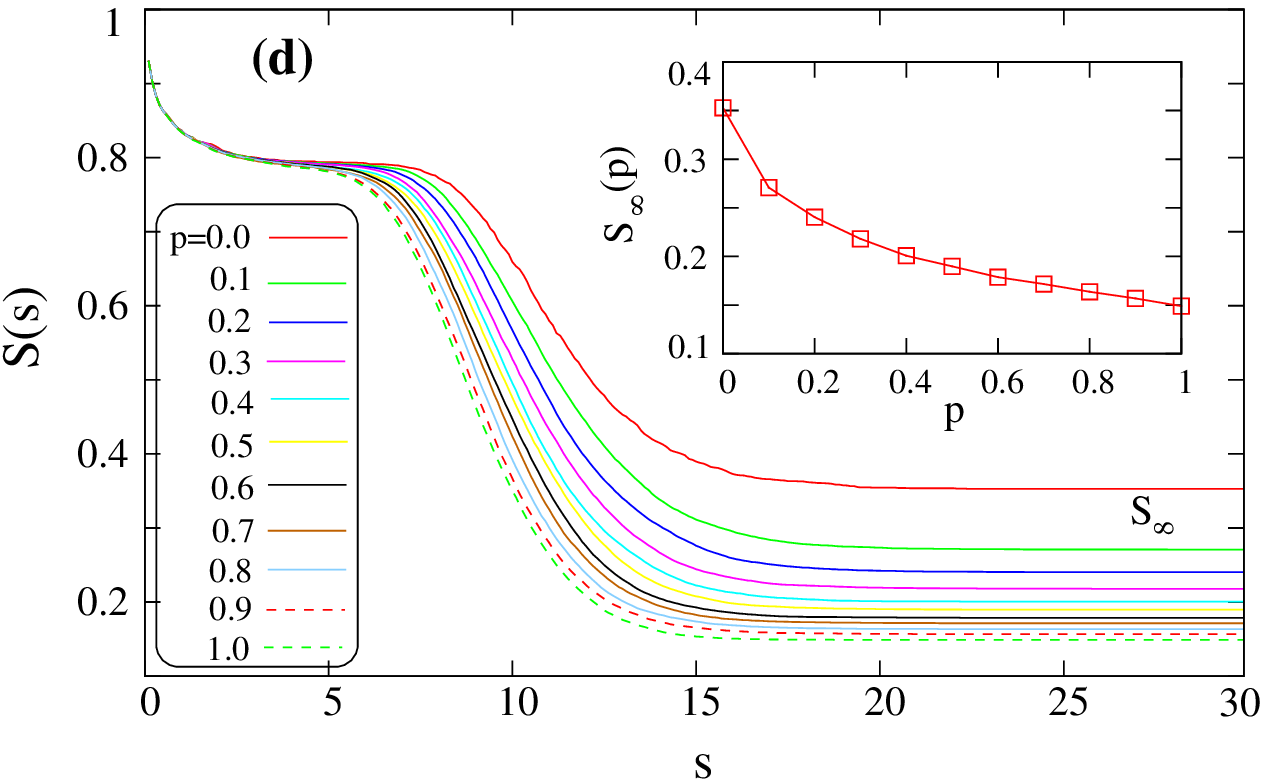}}\\
\end{tabular}
\end{center}
\caption{ (Color online) The plots of $S(s)$  as a funwction of reduced time $s$, for increasing values of wiring probabilities $p$. The  $1$-cycle scheme is shown in (a) for a $1$-dimensional ring and (c) for a $2$-dimensional square lattice. The $5$-cycle scheme is shown in (b) for a $1$-dimensional ring and (d)  for a $2$-dimensional square lattice.  The insets in all the figures show the asymptotic survival ratios $S_{\infty}$ as a function of the probability $p$. Here, the system size for the $1$-d ring is  $2000$ and for the $2$-d square lattice  it is $50 \times 50$ - all our data is averaged  over $10$ random network configurations.\label{one}}
\end{figure}

\begin{figure}[!t]
\begin{center}
\resizebox{90mm}{75mm}{\includegraphics{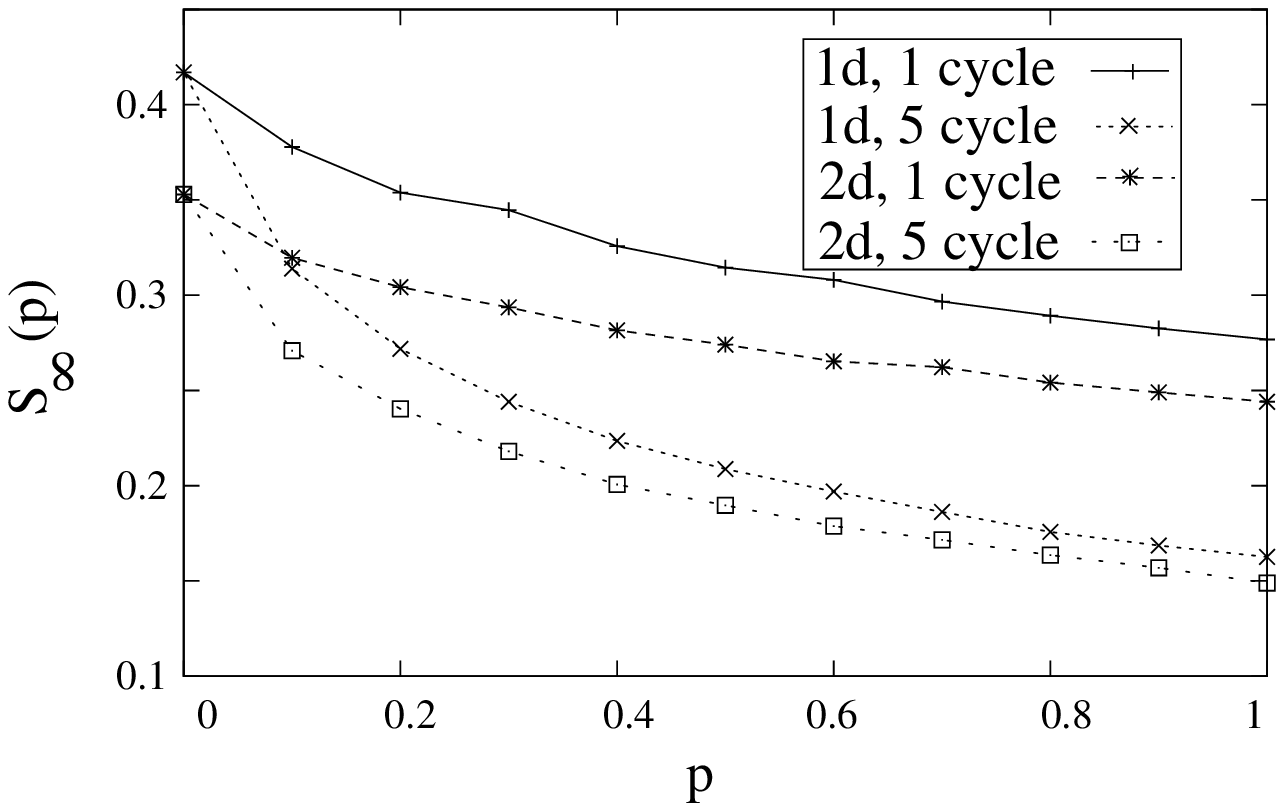}}\\
\end{center}
\caption{The plot of asymptotic survival ratios $S_{\infty} (s)$ as a function of the probabilities $p$ for $1$- and $5$-cycle schemes in $1$ dimension and $2$ dimensions. The error bars for all the graphs do not exceed $0.003$ and are smaller than the plot symbols.\label{asymp_all}}
\end{figure}
\subsection{One-dimensional ring and two-dimensional square lattices}

First, we consider a regular one-dimensional ring lattice of size $N=2000$. To start with, the clusters located on the lattice sites evolve according to  Eqn. \ref{finite_model},  where the interactions are  with nearest neighbours only. Next, we modify  the lattice by adding new links between sites chosen randomly with an associated probability $p \in [0,1]$. For $p=0$, the network is undisturbed and therefore ordered, while for $p=1$, the network becomes completely random. 

In our first scheme, we add links probabilistically starting with site $i=1$ and end with $i=N$, only once: we call this the $1$-cycle scheme.  The survival ratios of clusters as a function of reduced time $s$ for different values of wiring probability $p$ are presented in Fig. \ref{one}(a). Consider the $p=0$ case, which corresponds to a regular lattice; here the survival ratio $S(s)$ shows two stages,
Stage I  and Stage II, in its evolution. For all values of $0 < p \le 1$,  the existence of these well-separated Stages I and  II is also observed. There is a noticeable fall in the survivor ratio  as $p$ is increased, though; this is clearly visible in  the  asymptotic values $S_{\infty}(p)$ plotted with respect to $p$ in the inset of Fig. \ref{one}(a). As the probability $p$ increases, the number of links increases leading to more interaction and competition between the clusters and hence  a  decrease in the number of survivors. As expected,  this behaviour interpolates between the two characteristic behaviours  relevant to the regular lattice and  mean field scenarios.

Next we implement the $5$-cycle scheme, where the rewiring is done five times. Figure \ref{one}(b) shows the survivor ratio $S(s)$ as a function of $s$ for this scheme. We can clearly see the decrease in the number of survivors at all times, for increasing values of $p$. The asymptotic survival ratios $S_{\infty}$ for the $1$-cycle scheme and $5$-cycle scheme are shown in the insets of Fig. \ref{one}(a) and (b), respectively, where the monotonic decrease of $S_{\infty}$ with increasing $p$ is very clear for both cases. In addition, for all values of $p$, the survivor ratio in the $5$-cycle scheme is  consistently smaller than in the $1$-cycle scheme.

The same procedure for  a two-dimensional square lattice of size $50\times50$  is followed, and  survival ratios obtained as a
 function of $s$ using the $1$-cycle scheme and $5$-cycle scheme (see Figs. \ref{one}(c)-(d)). The asymptotic survivor ratios
for these two cases are shown in the insets of Figs. \ref{one}(c)-(d); they follow a decreasing trend with increasing $p$, similar to the one-dimensional case. We plot the asymptotic survival ratios of the $1$-cycle and $5$-cycle schemes in $1$-dimensional and $2$-dimensional lattices in Fig. \ref{asymp_all} for comparison.  Notice that the trends in  the plots for the $5$-cycle scheme are far smoother  than those for the 
corresponding $1$-cycle plots. This is expected  because of better statistics achieved in the $5$-cycle scheme.

\section{Damage spreading results\label{sec_dam}}
In this section, we  measure the  sensitivity of the system to slight perturbations in terms of damage spreading. The concept of damage spreading was introduced
in the context of biological systems \cite{kauff} and was widely used in the study of the dynamics of statistical mechanical systems \cite{derrida, stanley}. 
A system is said to have damage spreading if the {\it Hamming distance}  between two replicas, dictated by the same evolution equations, under slightly different initial conditions, grow with time. This concept has also been used to study phase transitions in  random networks  \cite{luque}.
However, our definition of damage spreading is different from the usual  definition based on the Hamming distance, used by workers on Ising models; we will elaborate on this below.

\subsection{Sensitivity to small perturbations}
\begin{figure}[!b]
\begin{center}
\hspace{-0.4cm}\resizebox{90mm}{70mm}{\includegraphics{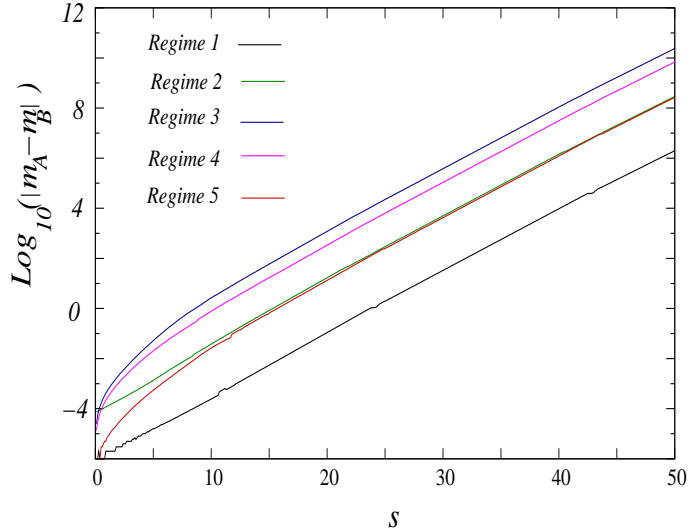}}\\
\end{center}
\caption{(Color online) The absolute mass difference $|m_A - m_B|$ of two mass configurations $\Omega_A$ and $\Omega_B$, as a function of the reduced time $s$. The case shown here is for a regular $1$-dimensional regular ring lattice of size $1000$. \label{sensitivity}}
\end{figure}

Again, we consider a $1$-dimensional ring lattice with only nearest-neighbour interactions. An exponential distribution of initial masses of clusters is put on the lattice, as before. Starting with two identical initial mass configurations $\Omega_A$ and $\Omega_B$, we  swap two randomly chosen clusters in the configuration $\Omega_B$.
We recall from the discussion above that the evolution  of  clusters in a finite-dimensional lattice  consists of two stages, fast and slow, and  that an  initial  cluster can be a survivor at the end of Stage I only if its  mass $X_i>X_{\star}$.  Therefore, while perturbing the initial mass configuration $\Omega_B$, we consider a variety of regimes that takes into account all possible combinations of   clusters  that are swapped.  Table \ref{regimes} summarizes all such combinations of $X_i$ and $X_j$, scaled with respect to the threshold square mass $X_{\star}$. Next, we evolve the two configurations $\Omega_A$ and $\Omega_B$ in time and find the absolute difference of the \textit{total mass} of the two  configurations, $|m_A - m_B|$, with time. Figure \ref{sensitivity} shows the difference $|m_A - m_B|$ for all five regimes shown in table \ref{regimes}. Note 
that irrespective of the regime, the quantity $|m_A - m_B|$ diverges with time.  We conclude therefore that the system is very sensitive to the initial masses of the clusters, even  for the regular lattice, where \textit{no} random links are added.
\begin{table}
\begin{center}
\begin{tabular}{lcr}
\hline
\hline
Regime&$X_i$&$X_j$\\
\hline
Regime 1 & $<< X_{\star}$ & $<<X_{\star}$\\
Regime 2& $\leq X_{\star}$ & $<<X_{\star}$\\
Regime 3& $\leq X_{\star}$ & $>X_{\star}$\\
Regime 4& $\geq X_{\star}$ & $>X_{\star}$\\
Regime 5& $>X_{\star}$&$>X_{\star}$($ X_i \sim X_j$)\\
\hline
\end{tabular}
\end{center}
\caption{\label{regimes} The five regimes  defined here are with 
respect to the threshold  mass $X_{\star}$. The clusters 
located at sites $i$ and $j$ are swapped in the configuration $\Omega_B$. }

\end{table}

We now summarize the above results. First, even for the regular square lattice, we have demonstrated the complexity of the dynamics:  it is impossible to predict the fate of a randomly chosen cluster unless its initial mass is less than the critical mass $X_{\star}$. In Sec. \ref{sec_compnet} we have started our study of the model on a network structure by adding new links (static wiring) the square lattice; we have shown that as the probability of random connections $p$ increases, the asymptotic number of  survivors decreases monotonically. This decrease, attributable to the death of clusters with marginal masses when their contacts increase in number,  matches with the results of the mean field scenario \cite{luck_am}. From a damage spreading point of view, we also see that
the system displays great sensitivity to the distribution of mass within it; the average mass difference between a perturbed and parent cluster diverges in time, irrespective of the masses of clusters that are swapped. 

Given all of this, we might guess that it would be even more difficult
 to predict the future of a randomly chosen cluster (again, for $X_i>X_{\star}$) when the model is solved on a network. In the next section, we attempt this in order to address the following question -  can selective networking be used to change the fate of a given cluster? 
 
\section{Networking strategies\label{networksmall}}

Since the central issue in this model is the survival of clusters against the competition,  the most interesting use of networks would be in the finding of
 a smart networking strategy which is able to change the fate of a cluster. This can be accomplished by  making a marginal survivor into a strong one, and more dramatically, by seeing if a  \textit{dying cluster can be revived to life}.

We systematically investigate  the effect of adding a finite number of nonlocal connections to a chosen central cluster. In Sec. \ref{regular}, it was shown that  the growth or decay of a cluster is solely dictated by its relative rate of change versus the cumulative rate of change of its neighbours. The key to better survival should
therefore lie in choosing appropriately decaying non-local clusters to network with; we will show
 in the following, that the fate of a cluster
can indeed be modified if this is done.

Given the complexity of the problem, we cannot categorize the clusters based on their rates of growth; the
 most viable option is a categorization of the clusters based on their initial masses. We  divide the nonlocal connections into two classes: viz., class A comprises  eventual non-survivors ($X<X_{\star}$), while class B comprises would-be survivors ($X>X_{\star}$) ). In the next subsection, we look at the outcome of networking with members of class A. 

\subsection{Nonlocal connections with  eventual non-survivors ($X_i < X_{\star}$)}
Recall that non-survivors ($<X_{\star}$) die very early during  Stage I. In connecting such
clusters to a given cluster with $X>X_{\star}$, we can be sure that they will never be able to
compete with it, much less kill it. Could such `harmless' clusters, nevertheless be useful in enhancing the survival chances of clusters with  $X>X_{\star}$?

\begin{figure}[!t]
\begin{center}
\begin{tabular}{cccc}
\hspace{-1cm}(a)&
\hspace{-0.4cm}\resizebox{50mm}{50mm}{\includegraphics{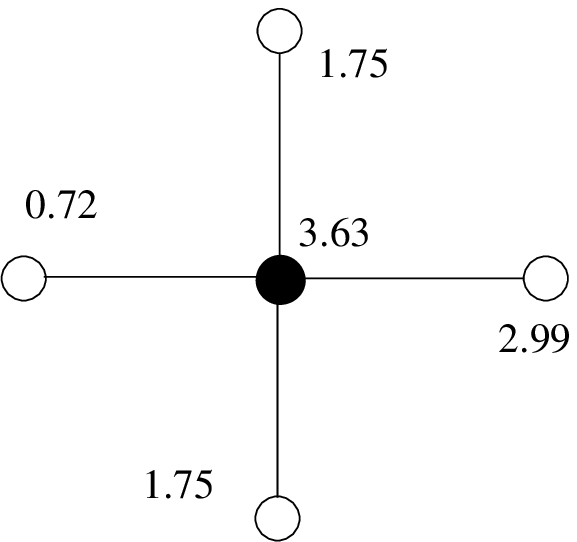}}&
\hspace{0cm}(b)&
\hspace{0.0cm}\resizebox{70mm}{60mm}{\includegraphics{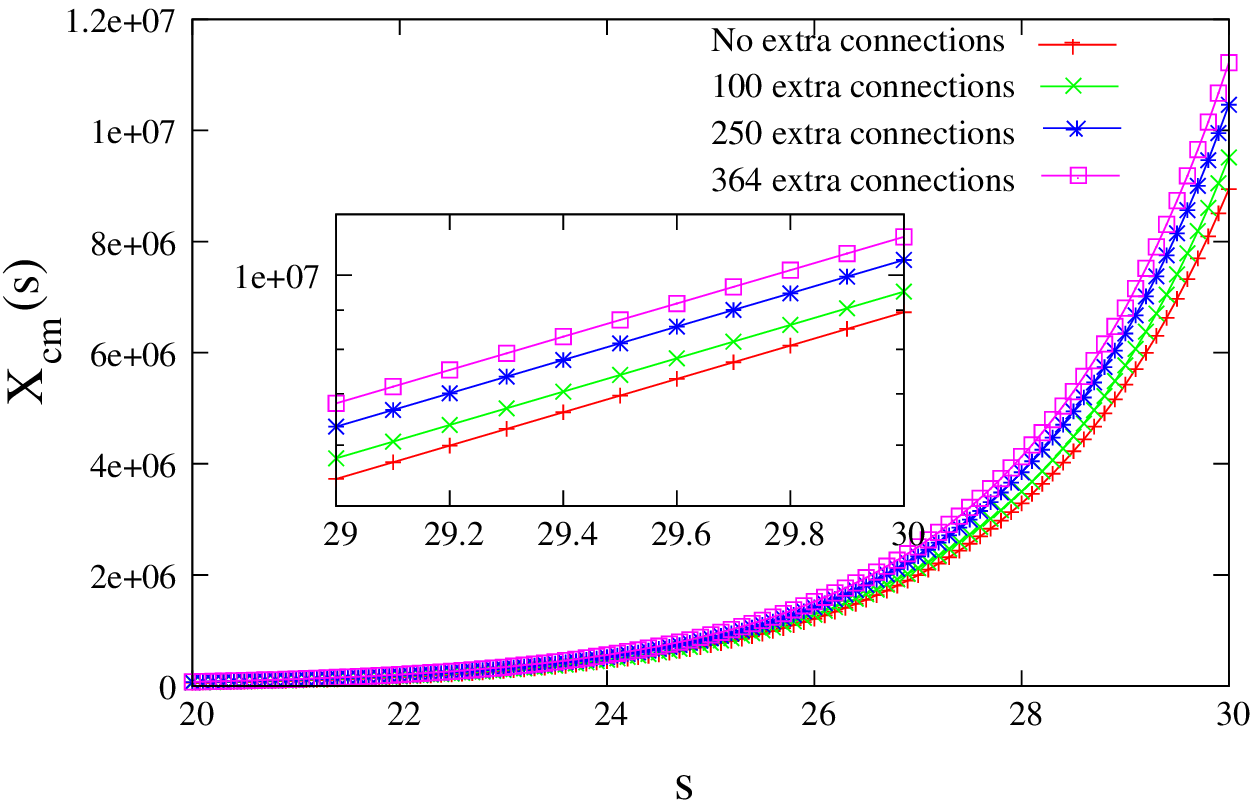}}\\
\end{tabular}

\end{center}
\caption{ (Color online) (a) The central cluster here is a survivor in its original configuration. (b) The asymptotic mass of the  central cluster diverges with increasing connectivity with  eventual non-survivors. \label{nonsurvive}}
\end{figure}

Let us consider a central cluster that will be an eventual survivor on its own in a given neighbourhood; one such case is shown in Fig. \ref{nonsurvive}. We now let the central cluster  network with eventual non-survivors from all over the lattice, and record its growth as a function of the number of the clusters it networks with; the results are shown in Fig. \ref{nonsurvive} (b). When the neighbours of the  central cluster die, their contribution goes to zero (see Eqn. 3) and the solution to the resulting  first order linear differential is therefore an exponential as shown in Fig. \ref{nonsurvive} (b).
It is evident that there is a marked increase in the mass of   the central cluster as  more small clusters are connected to it; since  none of these will  survive past Stage I, their cumulative growth rate
will be negative, leading to an increase in the  rate of growth of the chosen cluster. This makes it a better (more massive) survivor 
 asymptotically. As the number of nonsurviving contacts is increased, the mass of the chosen cluster increases (Fig. \ref{nonsurvive}).

\begin{figure}[!b]
\begin{center}
\begin{tabular}{cc}
\hspace{-1cm}(a)&
\hspace{-0.4cm}\resizebox{90mm}{70mm}{\includegraphics{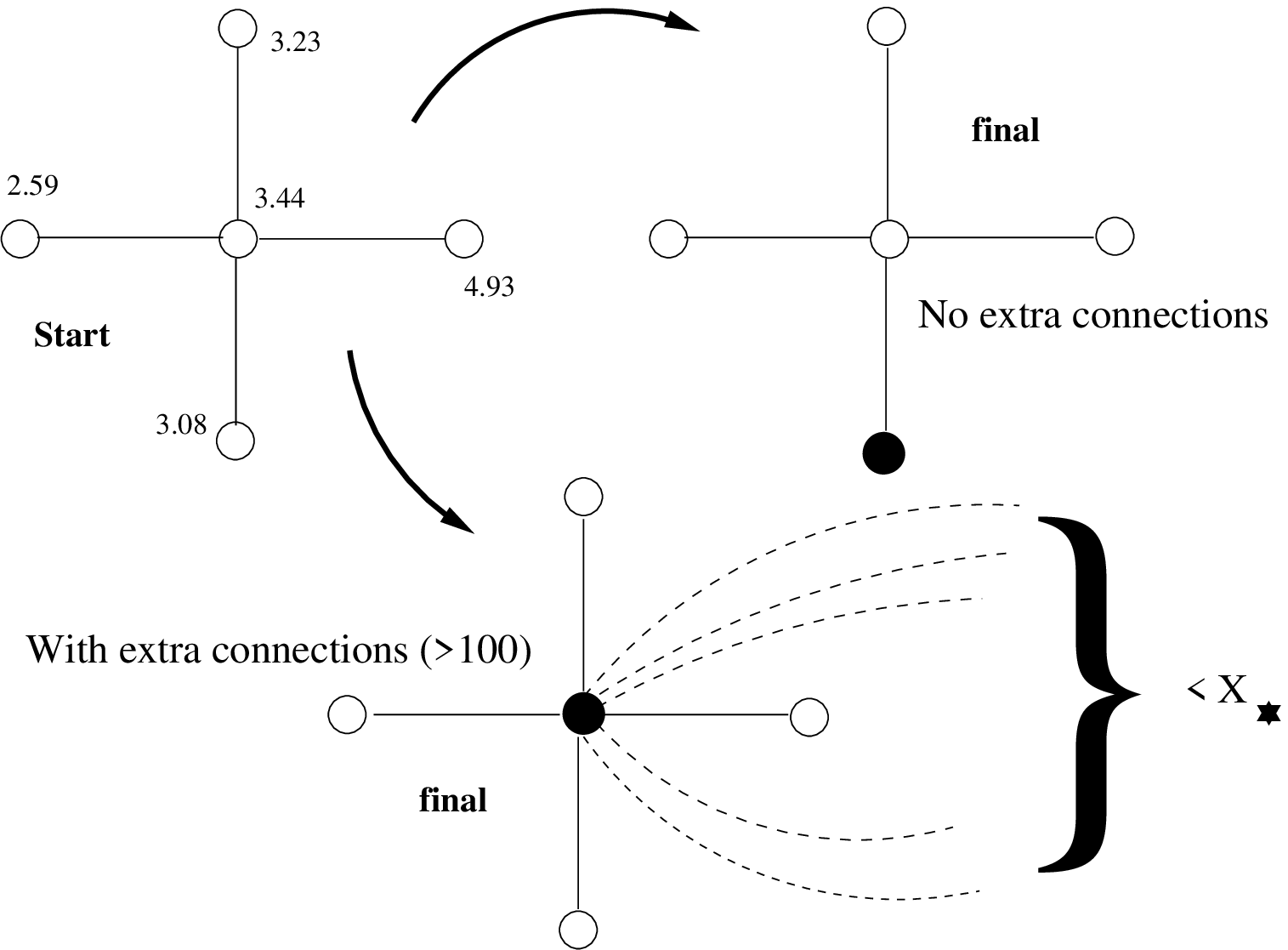}}\\
\hspace{-1cm}(b)&
\hspace{-0.4cm}\resizebox{90mm}{60mm}{\includegraphics{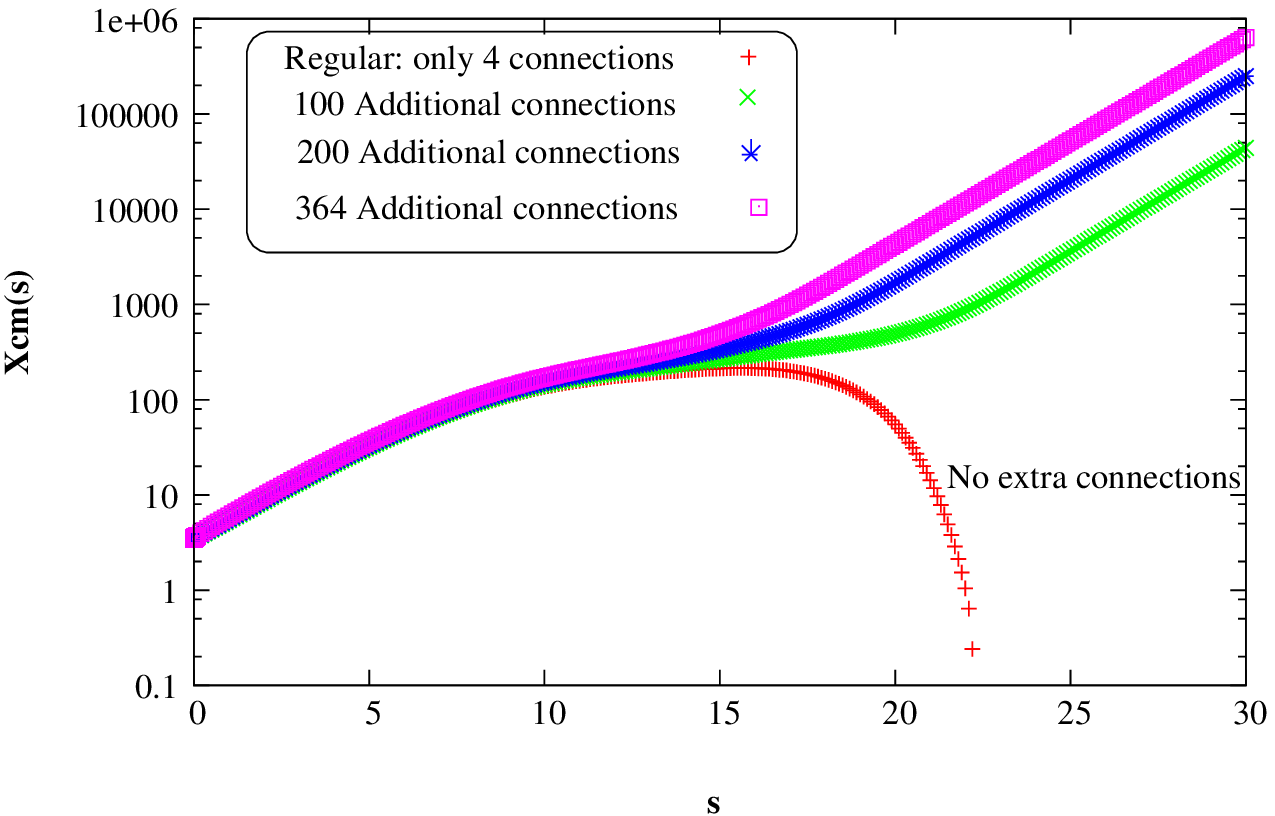}}\\
\end{tabular}

\end{center}
\caption{ (Color online) (a)  The central cluster dies in its original configuration without 
extra connections. On addition of more and more clusters it becomes a survivor. (b) A crossover is seen here as the 
central cluster becomes a survivor from being a nonsurvivor, with the increased number of connections to non-surviving clusters ($< X_{\star}$).  \label{givelife}}
\end{figure}

We now use this observation to try to change the fate of  a   dying cluster, and  turn it into  a survivor. Although the nonlinear, many-body nature of the problem makes it rather difficult to construct such cases, we have been able to network  an erstwhile dying cluster with suitable nonlocal contacts, and returning it to life. Figure \ref{givelife} shows our results:   the central cluster would eventually have died in its original environment, but on  adding $100$ small clusters  (whose mass $< X_{\star}$) it manages  to live. Further additions, e.g.  $200$  or  $364$ clusters evidently make it a better survivor (Fig. \ref{givelife} (b)).

\subsection{Networking with would-be survivors $X_i > X_{\star}$}
Here we are dealing with a far more complex problem than the case discussed in earlier subsection. 
When we choose clusters with masses larger than $X_{\star}$, their lifespan will exceed Stage I; thus during Stage I they will have a positive growth rate. Depending on their individual environments, they may survive through
Stage II - thereby having a positive growth -  or die as a result of a negative growth rate. Networking with such clusters is therefore a rather delicate matter.

\begin{figure}[!t]
\begin{center}
\resizebox{120mm}{100mm}{\includegraphics{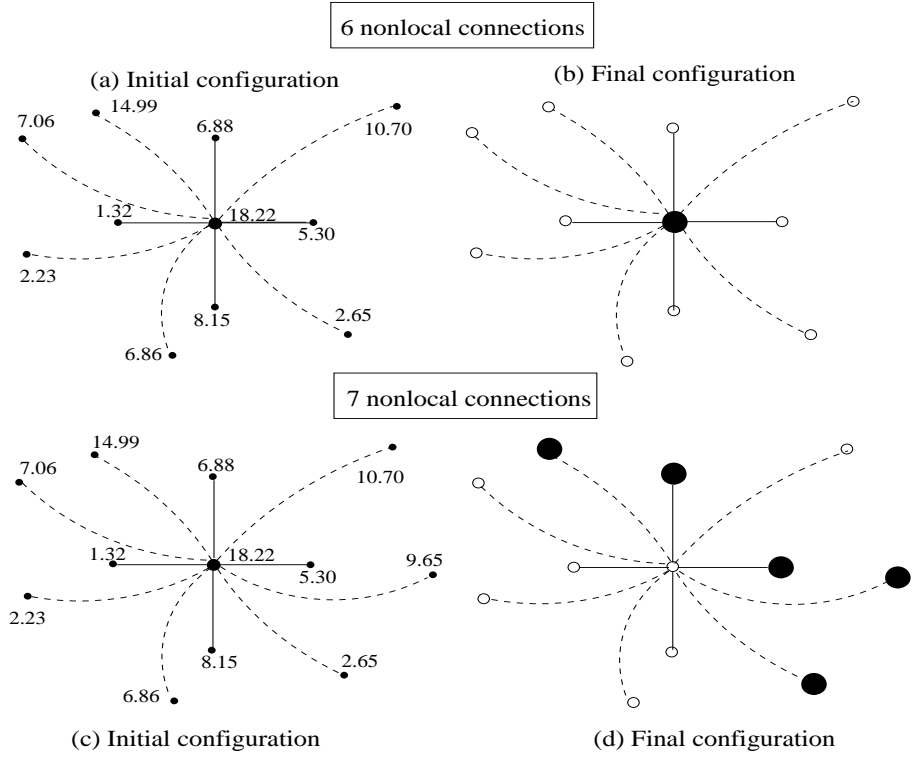}}\\
\end{center}
\caption{The central cluster is networked with would-be clusters ($X_i>X_{\star}$) that are less massive than the central cluster ($X_{cm}$).   (a) The initial configuration, where the central cluster has $6$ nonlocal connections, (b) the asymptotic state with the only survivor being the central cluster. (c) With the addition of one more nonlocal cluster to the existing configuration  (d) the central cluster dies. In (b) and (d), the open circle represent nonsurvivors and dark circles represent asymptotic survivors. \label{schem636}}
\end{figure}

First, we consider nonlocal connections with  would-be survivors ($X_i > X_{\star}$) which are
less massive than our chosen cluster ($X_i < X_{cm}$). Such would-be survivors will live beyond Stage I and will grow with a positive
rate.  This is shown clearly in Fig. \ref{freemasses} where clusters with $X_i < X_{\star}$ die in finite time and clusters with $X_i > X_{\star}$ grow beyond stage I.  From a  mean field perspective,  we would therefore expect 
to see  a decrease in the chances of survival of the chosen cluster  as it networks with more and more would-be survivors.

\begin{figure}[!b]
\begin{center}
\resizebox{120mm}{90mm}{\includegraphics{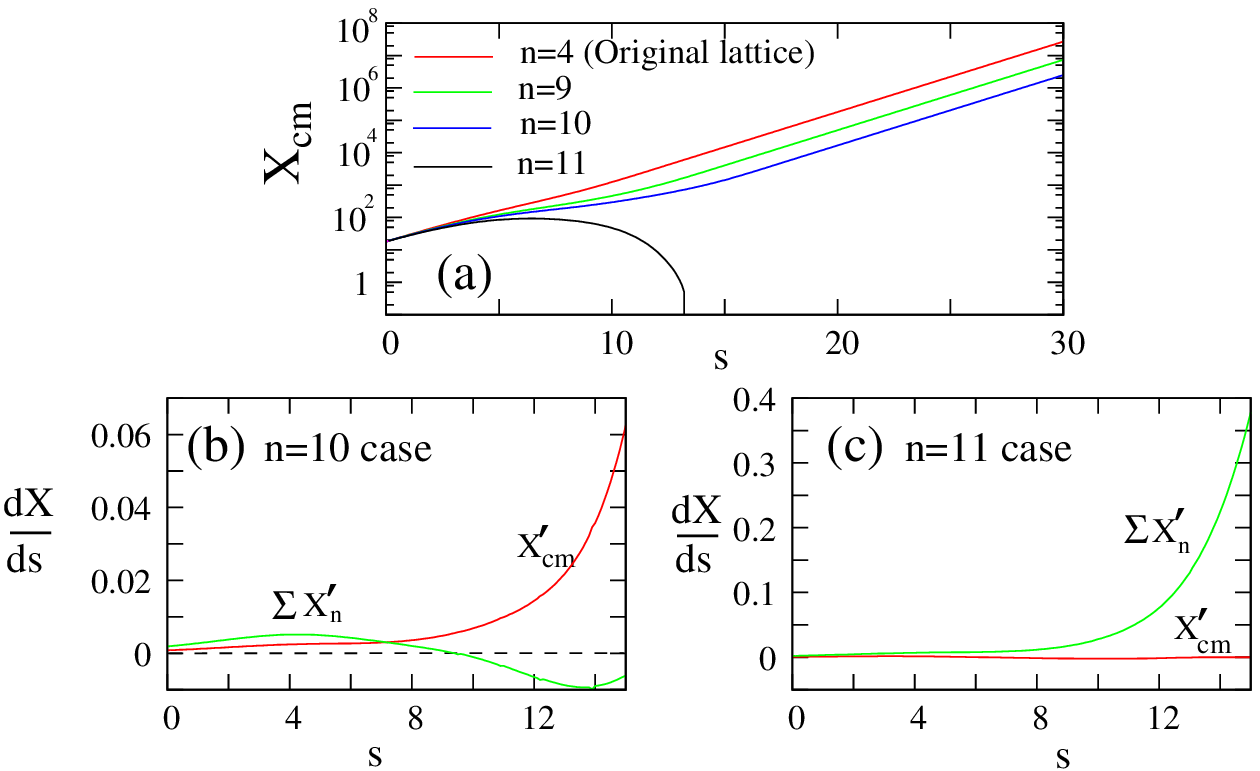}}\\
\end{center}
\caption{ (Color online)The central cluster of a configuration given in Fig. \ref{schem636}(a) is networked with clusters whose $X_n > X_{\star}$. In (a) the growth of $X_{cm}$ with an increasing number of connections is plotted -  $n=4$ corresponds to a regular lattice. There is a crossover seen when the number of connections increases from  $n=10$ to $n=11$; for larger $n$ values, the central site dies. This observation is supported by the  rates of growth of $X_{cm}$ and its neighbours $X_{n}$   (b) when $n=10$ and (c) when $n=11$. \label{636}}
\end{figure}

 Figure \ref{schem636} shows a sample scenario, where the central cluster is connected with nonlocal clusters which are not only would-be survivors  ($X_i > X_{\star}$) , but also less massive than the central cluster. In Fig. \ref{schem636} (a) the central cluster networks with 6 would-be survivors and is able to survive in the asymptotic state (Fig. \ref{schem636} (b) ). On the other hand, adding one more would-be survivor (Fig. \ref{schem636} (c)) to the existing network of the central cluster  kills it in the asymptotic state (Fig. \ref{schem636} (d)). We also note that, this change in the fate of the central cluster - from being a survivor to being a dying cluster, due to the additional connection - also changes the fate some of the  clusters linked to the central cluster (contrast Figs. \ref{schem636} (b) and (d)).

To understand the dynamics leading to the observed behaviour we look at the rates of growth of the central cluster
and its neighbours. A sample result corresponding to the cases shown in Fig. \ref{schem636} is discussed in Fig. \ref{636}.
Here, we see that  increasing the number of nonlocal connections  with would-be 
survivors leads to a fall in the absolute value of $X_{cm}$ as well as its rate of growth $X'_{cm}$. Beyond
 a certain number of networked contacts, the  central cluster begins to decay, and eventually dies. This crossover from life to death happens when the cumulative rate  of growth of the neighbours $\Sigma X'_{i,j}$  
 is larger than that of the central cluster $X'_{cm}$. Unfortunately the intricate many-body nature of this problem precludes a prediction of when such crossovers might occur in general. We will have more to say on this issue in the next section.

\begin{figure}[!t]
\begin{center}
\resizebox{120mm}{100mm}{\includegraphics{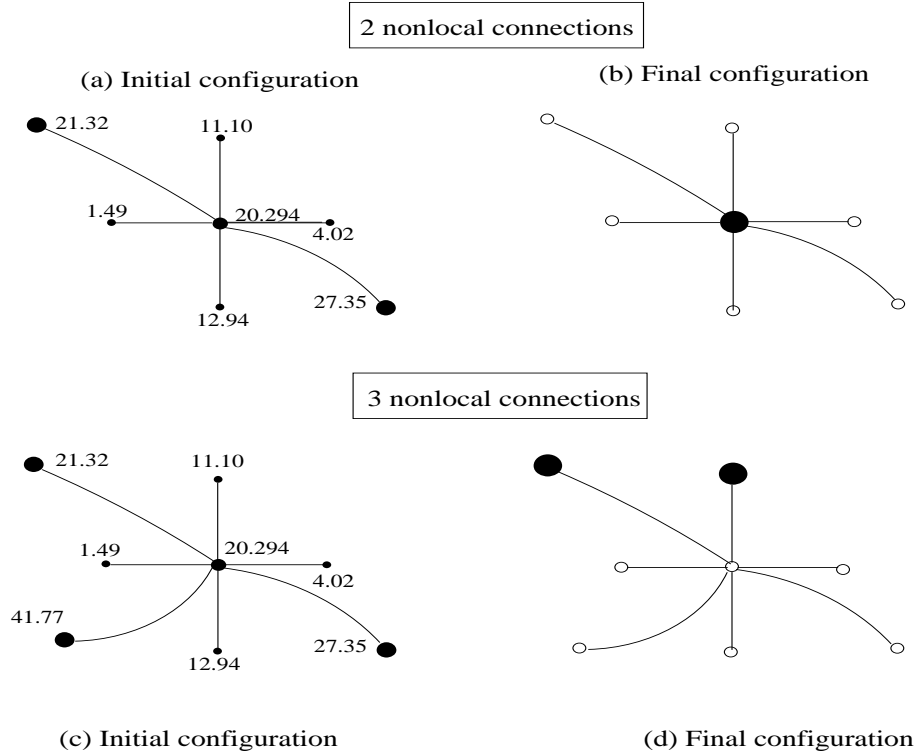}}\\
\end{center}
\caption{The central cluster is networked with would-be clusters ($X_i>X_{\star}$) that are more massive than the central cluster ($X_{cm}$).   (a) The initial configuration with $2$ nonlocal connections, (b) and its asymptotic state; the only survivor here is the central cluster. (c) With the addition of one more nonlocal cluster to the existing configuration (d) the central cluster dies. In (b) and (d), the open circle represent nonsurvivors and dark circles represent asymptotic survivors.\label{schem1054}}
\end{figure}

\begin{figure}[!b]
\begin{center}
\resizebox{120mm}{90mm}{\includegraphics{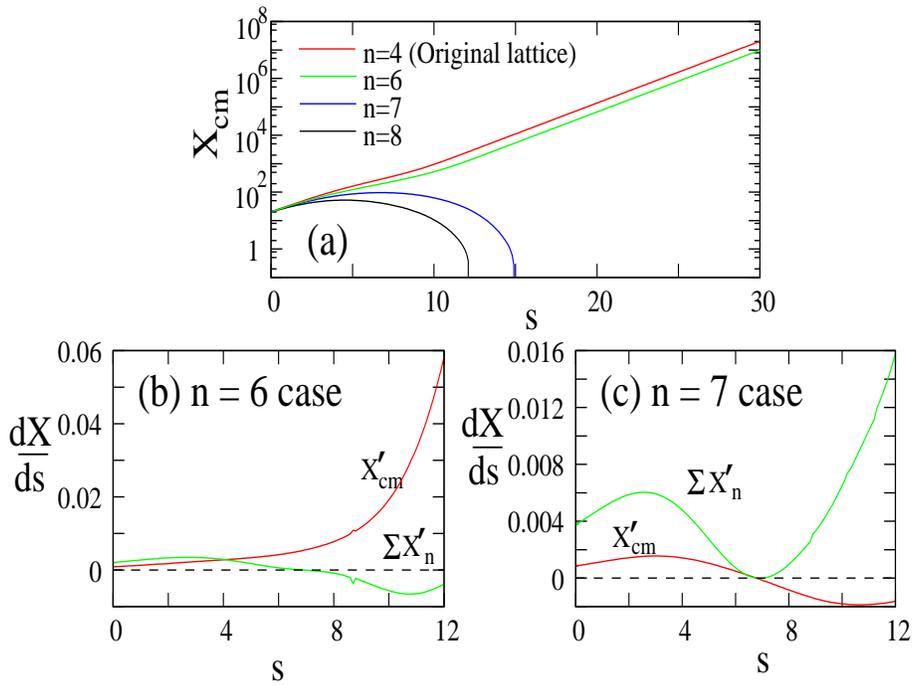}}\\
\end{center}
\caption{ (Color online) The central cluster of a configuration given in Fig. \ref{schem1054}(b) is connected with nonlocal clusters with $X_n > X_{cm}$. In (a) the growth of $X_{cm}$ with an increasing number of connections is plotted -  $n=4$ corresponds to a regular lattice. There is a crossover seen when the number of connections increases from  $n=6$ to $n=7$; for larger $n$ values, the central site dies. This observation is supported by the  rates of growth of $X_{cm}$ and its neighbours $X_{n}$   (b) when $n=6$ and (c) when $n=7$. \label{1054}}
\end{figure}

 Finally, in the case where a given  cluster networks with larger would-be survivors ($X_i > X_{\star}$ and $X_i > X_{cm}$), one would expect a speedier death. One such sample scenario is depicted in Fig. \ref{schem1054} and the corresponding rates of evolution of the clusters in Fig. \ref{1054}.  We notice that in its original configuration the central cluster (with four neighbours, ($n=4$)) is a survivor. As we increase the number of networked connections, its growth gets stunted;  there is a 
substantial fall for 2 extra links ($n=6$ in Fig. \ref{1054}). Adding one more link ($n=7$) does the final damage; the central cluster dies.  The rates shown in Fig. \ref{1054} (b) and (c) for $n=6$ connections and $n=7$ connections
vividly capture the competition for survival, leading to life in one case and death in the other.

As expected, we observe that  fewer connections (here, $n=7$) are needed, compared to the earlier case with smaller would-be survivors ($n=11$), to kill the chosen cluster. In closing, we should of course emphasise that these results are sample results chosen to illustrate a qualitative point: the choice of different nonlocal sites to network with would certainly modify our $n$ values.

\section{Universality in the distribution of survivors - A clue to survival against the odds? \label{univ}}

The results of the previous section indicate that if a cluster wants to change its fate, it should always network with non-survivors; however, it seems intuitively reasonable to suppose that it might also be able to do this if it networks with would-be survivors, provided that they are much smaller than itself.
Finding such cases is possible ex post facto; their prediction is next to impossible given the complexity of the problem. 

An interesting  case to consider  might be one where, against the odds, the most massive cluster in a given neighbourhood dies marginally, because of  the unexpected growth of its erstwhile smaller neighbours. If, now, it is able to network with potential survivors much smaller than itself (who, as a result of their own dynamics, would eventually perish in their own neighbourhoods), this would be a promising way of selectively networking itself to life. Similar issues are associated with survivors who survive against the odds; they could be killed off by suitable networking.

A systematic way of exploring this question is to look at the statistics associated with some of these rare events. We look first at 
the four  immediate neighbours  of a given cluster, and consider their pairwise interactions with it. Clearly, had such a pair been isolated, the larger cluster would have won \cite{luck_am}. However, many-body interactions in the lattice mean that this is not always true. We therefore ask the question:   \textit{what is the proportion of}  cases  where the larger cluster wins?  
\begin{figure}[!t]
\begin{center}
\resizebox{120mm}{90mm}{\includegraphics{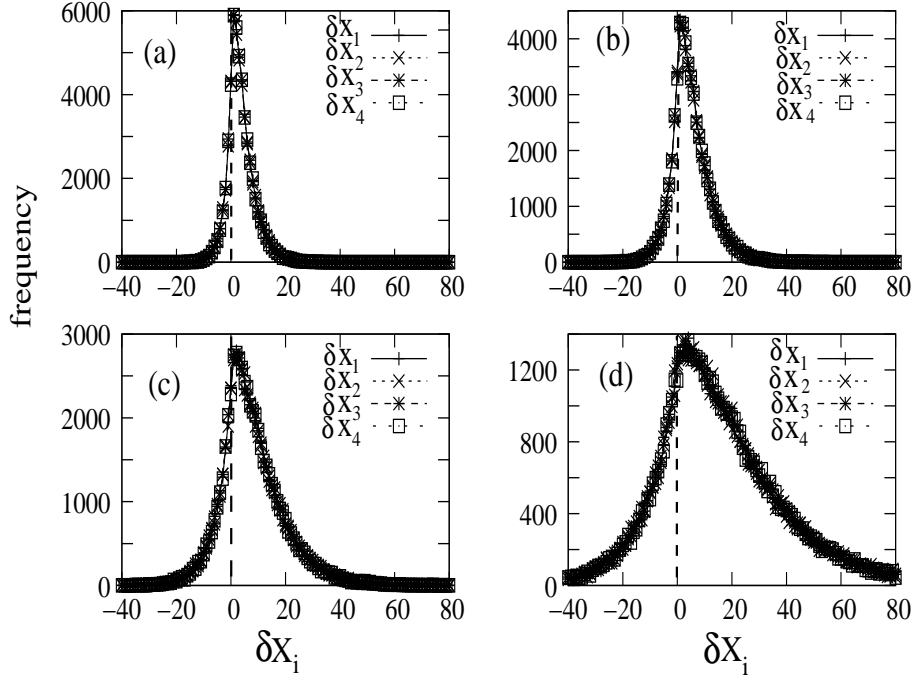}}\\
\end{center}
\caption{The plots show the distribution of pairwise mass differences between  survivors and their four neighbours. They are obtained for  exponential distributions of initial masses with  different mean values   $1/\mu$, where $\mu = -\log(S_1)/X_{\star}$.  The plots are for (a) $1/\mu = 3.92$ ($S_1 = 0.6$) (b) 5.607  ($S_1 = 0.7$) (c) 8.963  ($S_1 = 0.8$) and (d) 18.982  ($S_1 = 0.9$). The system size is $400 \times 400$. \label{univ_fig}}
\end{figure}

\begin{figure}[!t]
\begin{center}
\resizebox{80mm}{60mm}{\includegraphics{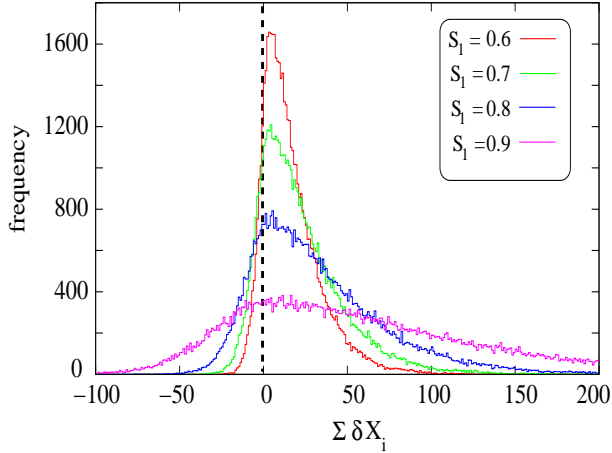}}\\
\end{center}
\caption{(Color online) The plots show the distribution of mass differences between survivors and all of their four neighbours. They are obtained for  exponential distributions of initial clusters with different mean values  $1/\mu$, where $\mu = -\log(S_1)/X_{\star}$.  The plot in  `red' represents the case for  $1/\mu = 3.92$ ($S_1 = 0.6$), `green' represents $1/\mu = 5.607$  ($S_1 = 0.7$), `blue' represents $1/\mu = 8.963$  ($S_1 = 0.8$), and  `pink' represents $1/\mu = 18.982$  ($S_1 = 0.9$). The system size is $400 \times 400$. \label{univ_cuml}}
\end{figure}

\begin{figure}[!t]
\begin{center}
\resizebox{120mm}{90mm}{\includegraphics{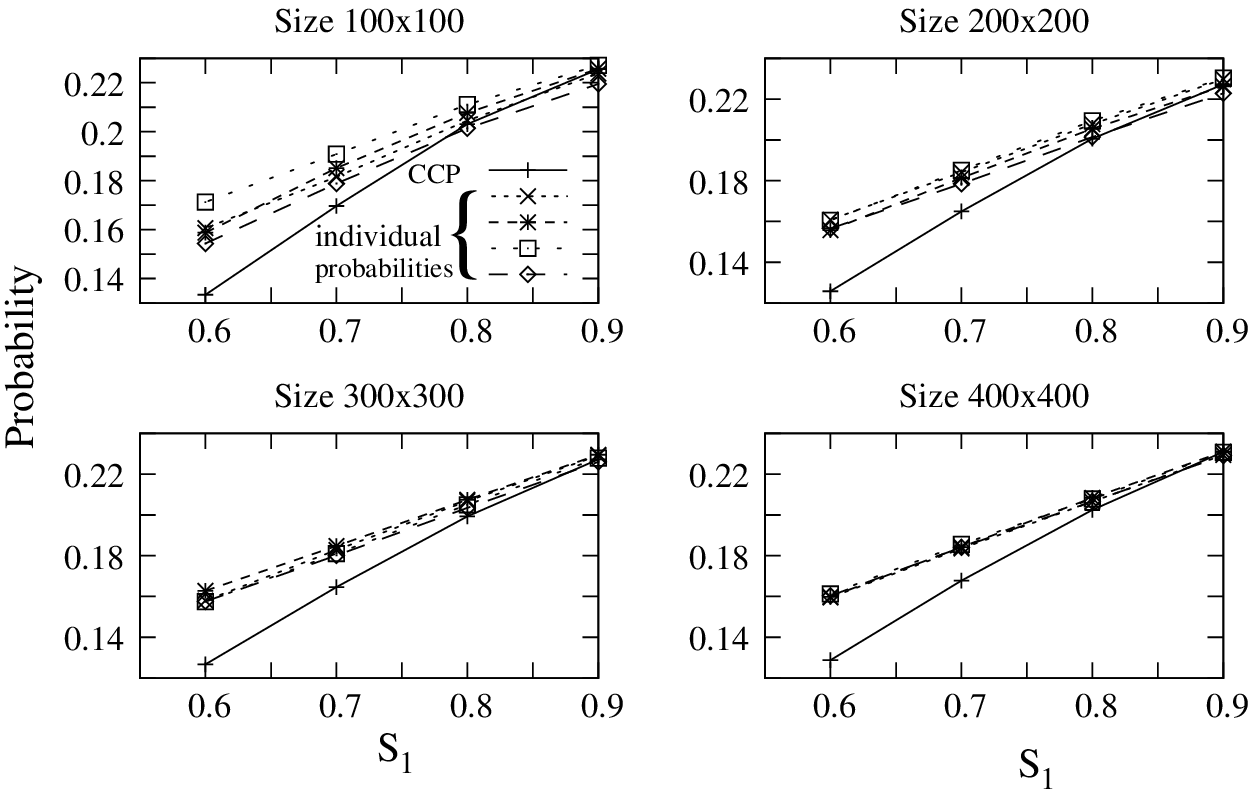}}\\
\end{center}
\caption{The plots show the  probability of finding a successful central cluster with a larger neighbouring cluster, to start with. The probability increases with increase in $S_1$ (refer to Figs. \ref{univ_fig} and \ref{univ_cuml}). While the solid line show the combined cumulative probabilities (CCP) (from Fig. \ref{univ_cuml}), the rest of the lines represent the individual cumulative probabilities (from Fig. \ref{univ_fig}). \label{all_prob}}
\end{figure}

Each survivor has four neighbours; we first calculate the probability distribution of the initial mass differences in a pairwise fashion between a survivor and each of its neighbours. The initial mass differences are given by $\delta X_i = X_{cm} - X_i$ ($i=1,2,3,4$)  corresponding to the four neighbours - right, left, bottom and top - of a survivor. The distribution of $\delta X_i$ for all the survivors is shown in Fig. \ref{univ_fig}. Here, a negative $\delta X_i$ means that the survivor  is smaller than its neighbour, and conversely for positive $\delta X_i$. All four  distributions corresponding to  four neighbouring pairs  overlap due to isotropy; the resulting distributions are universal functions of {\it mass differences}, depending  only on $\mu$ .

We also obtain the cumulative mass difference between survivors and all of their four neighbours  viz. $4 X_{cm} - \sum_{i=1}^4 X_i = \sum_{i=1}^4 \delta X_i$ (see Fig. \ref{univ_cuml}). The distributions of $\sum_{i=1}^4 \delta_i$  are plotted in Fig. \ref{univ_cuml} for different values of $\mu$. For a positive cumulative mass difference we know that the survivor has a larger initial mass  than the sum of its neighbours, matching our intuition based on the mean-field regime. The  negative side of the distribution is more interesting, comprising survivors whose initial masses are smaller than the sum of the initial masses of their immediate neighbours.

 Notice that  both the survivor-neighbour pair distribution (Fig. \ref{univ_fig}), and the survivor- all neighbours cumulative distribution (Fig. \ref{univ_cuml}) get broader with increasing $\mu$.  This is because increasing $\mu = - \log(S_1)/X_{\star}$ \cite{luck_am} increases the number of potential survivors $S_1$ beyond Stage 1.   In each case,  the fraction of area under the negative side of the survivor pair-distribution
 gives an estimate of  survivors against the odds -- an example of  some of the rare events alluded to at the beginning of this section.
   
  Figure \ref{all_prob} shows this fraction, both in terms of  individual survivor-pair distribution and cumulative distribution, as a function of the 
$\mu$ of the initial mass distribution, for different system sizes. 
There are more survivors, hence more survivors
\textit{against} the odds, leading to an increase in the fraction plotted on the y-axis of
Fig. \ref{all_prob} for both  distributions. For the largest
system size, there is full isotropy in the pairwise distributions; the probability of finding a survivor
against the odds
 is now  seen to be a {\it regular and
universal function} of  $\mu$ in both pairwise and cumulative cases, relying only on mass differences
rather than on masses. Finally, the cumulative distribution gives a more stringent survival criterion than the pairwise one, as is to be expected from the global nature of the dynamics.

A major conclusion to be drawn from Fig. \ref{all_prob} is the following: there are
clusters which \textit{die} against the odds (clusters which are more massive than the survivor
in either a pairwise or cumulative sense). Recalling again that the distributions plotted
are independent of mass, it should therefore be possible to pick dying clusters whose mass
was greater than threshold and to network them with their peers so that they can survive. This will
be the subject of future work.

\section{Discussion and Conclusion \label{discuss}}

In this paper we have investigated the cluster growth model of \cite{luck_am} on a square lattice. Our main emphasis has been on stochastic networking strategies: long-range connections are introduced with  probability $0<p<1$ to existing lattice sites.

 We find first of all that the qualitative features of the networked system remain the same as that of the regular  case, in that the presence of two well-separated dynamical stages is retained; the glassy dynamics and metastable states of the earlier work \cite{luck_am} persist. However,  the number of survivors decreases as expected with increasing $p$, in accordance with a mean field perspective. Also, the system is very sensitive to the smallest changes in the initial mass configuration; the divergence is, interestingly,  exponential irrespective of the  magnitudes of the masses swapped. To quantify this,  we have used the concept of  damage spreading.

The central result of this paper is the use of smart networking strategies  to modify the fate of an arbitrary cluster. We find that it is safest to network with non-survivors; their decay and eventual death lead to the transformation of the destiny of a given site, from death to survival, or from survival to stronger survival. Networking with peers or with those of larger initial mass in general leads to weakening, and an almost inevitable death.

However, the above is not immutable:  the  probability distributions in the last section of the paper indicate an interesting universality of survival `against the odds'.  In future work, we will exploit this result to develop schemes for survival strategies; in complex systems of this type, it seems clear that the phrase `survival of the fittest' is more likely to be replaced by `survival of the best networked'.

\end{document}